# On a pipeline pressure drop model for nonideal, compressible, gas mixture flow with application to pipeline flow of natural gas-hydrogen blends

Jeremy Conner, Vasilios I. Manousiouthakis[*]

Department of Chemical and Biomolecular Engineering,

University of California at Los Angeles (UCLA), Los Angeles, California, 90095

* Corresponding Author, Email: vasilios@ucla.edu, Phone: (310) 206-0300, 5549 Boelter Hall, Box 951592, Los Angeles, CA, 90095-1592

## Abstract

This work presents a novel, dimensionless model that results in a dimensionless algebraic equation that can be used to quantify the pressure drop associated with the steady state, isothermal flow through a straight, horizontal pipeline, of a compressible gas mixture whose thermodynamic behavior is described for comparison purposes by ideal gas (IG) and nonideal gas generic cubic (GC) equation of state (EOS) models. A solution strategy, that uses the aforementioned dimensionless algebraic equation, is then presented that quantifies the pipeline pressure drop for hydrogen containing mixtures. Two case studies are presented to illustrate the pressure drop dependence on the hydrogen mole fraction of a binary, methane hydrogen mixture, and a natural gas hydrogen mixture with a real life natural gas composition containing eight species.

## Notation
**English Letters**

$A_p(m^2)$: Pipeline cross section considered constant

$a(T)\left((J \cdot m^3)/(mol^2)\right)$: Temperature dependent parameter of GCEOS

$b(m^3/mol)$: Parameter of GCEOS accounting for molecule finite size

$b^{IG}(m^3/mol) = \dfrac{b}{\Omega}$: Parameter accounting for molecule finite size, used in creating DIGEOS

$dl(m)$: Pipeline differential length

$d\hat{W}\left(\dfrac{m^2}{s^2}\right) = d\hat{W}\left(\dfrac{J}{kg}\right)$: Differential amount of work per unit mass provided to the gas mixture

$D_\beta \subset \mathbb{R}^+ \times \mathbb{R}^+$: Domain of $(q,\tilde{V})$ for DGCEOS

$D_p(m)$: Pipeline diameter considered constant

$D_P \subset \mathbb{R}^+ \times \mathbb{R}^+$: Domain of $(T,V)$ for GCEOS

$f(\cdot)$: Friction factor of nonideal gas mixture

$k(J/K) = 1.380649 \cdot 10^{-23}$: Boltzmann constant

$l(m)$ : Pipeline length running variable

$\tilde{l}(\cdot)$ : Dimensionless pipeline length running variable

$L_p(m)$ : Pipeline length

$\dot{m}\left(\dfrac{kg}{s}\right)$ : mass flow rate

$m_i(kg\,i)$ : mass of a single molecule of the ith species

$M\left(\dfrac{kg\,mix}{mol\,mix}\right)$ : Molar mass of nonideal gas mixture

$M_{NG}\left(\dfrac{kg\,NG}{mol\,NG}\right)$ : Molar mass of nonideal gas mixture

$M_i\left(\dfrac{kg\,i}{mol\,i}\right)$ : ith species' molar mass

$N_A(1/mol) = 6.02214076 \cdot 10^{23}$ : Avogadro number

$P(Pa)$ : Pressure of nonideal gas mixture

$P_{c,i}(Pa)$ : ith species' critical pressure

$q$ : dimensionless temperature of nonideal gas mixture

$r_{KI} = \Omega^2 \, r_{KI}^{IG}$ : dimensionless ratio of kinetic molar energy over internal molar energy for DGCEOS

$r_{KI}^{IG}$ : dimensionless ratio of kinetic molar energy over internal molar energy for DIGEOS

$r_{LD}$ : dimensionless ratio of pipeline diameter over pipeline length

$R(J/(mol \cdot K)) = 8.314$ : Universal Gas Constant

$Re(\cdot)$ : Reynolds number of gas mixture fluid flow

$R_h(m)$ : Pipeline hydraulic radius considered constant

$R_j$ : jth chemical reaction

$S_i$ : ith species

$T(K)$ : Temperature of nonideal gas mixture

$T_{c,i}(K)$ : ith species' critical temperature

$T_{r,i}(K)$ : ith species' reduced temperature

$T_i^*(\cdot)$ : ith species' dimensionless temperature normalized by Lennard Jones potential parameter

$v\left(\dfrac{m}{s}\right)$ : Gas mixture velocity

$v_0 \left( \dfrac{m}{s} \right)$: Gas mixture velocity at the pipeline entrance

$\tilde{v}(\cdot)$ Dimensionless velocity of gas mixture

$V \left( \dfrac{m^3 \, mix}{mol \, mix} \right)$: Molar volume of nonideal gas mixture

$V_0 \left( \dfrac{m^3 \, mix}{mol \, mix} \right)$: Molar volume of nonideal gas mixture at pipeline inlet

$\tilde{V}(\cdot) \triangleq \dfrac{V}{b}$: Dimensionless molar volume of gas mixture used in DGCEOS, normalized by the parameter $b$ accounting for molecule finite size.

$\tilde{V}^{IG} \triangleq \dfrac{V}{b^{IG}} = \Omega \tilde{V}$ Dimensionless molar volume of gas mixture used in DIGEOS, normalized by the parameter $b^{IG}$ accounting for molecule finite size.

$\tilde{V}_0(\cdot) \triangleq \dfrac{V_0}{b}$: Dimensionless molar volume of gas mixture normalized by GC EOS parameter $b$ accounting for molecule finite size at pipeline inlet.

$\tilde{V}_0^{IG}(\cdot) \triangleq \dfrac{V_0}{b^{IG}} = \Omega \tilde{V}_0(\cdot)$: Dimensionless molar volume of gas mixture normalized by parameter $b^{IG}$ accounting for molecule finite size at pipeline inlet.

$y_i \left( \dfrac{mol \, i}{mol \, mix} \right)$: ith species mole fraction

$Z_c$: compressibility factor at critical point

**Greek Letters**

$\alpha(T_{r,i}; \omega_i)(\cdot)$: Reduced temperature and acentric factor dependent parameter of GCEOS

$\beta \triangleq \dfrac{Pb}{RT}$: Dimensionless gas mixture pressure, normalized by temperature, used in DGCEOS

$\beta^{IG} \triangleq \dfrac{Pb^{IG}}{RT} = \dfrac{\beta}{\Omega}$ Dimensionless gas mixture pressure, normalized by temperature, used in DIGEOS

$\beta_0 \triangleq \dfrac{P_0 b}{RT}$: Dimensionless inlet gas mixture pressure, normalized by temperature, used in DGCEOS

$\beta_0^{IG} \triangleq \dfrac{P_0 b^{IG}}{RT} = \dfrac{\beta_0}{\Omega}$ Dimensionless inlet gas mixture pressure, normalized by temperature, used in DIGEOS

$\gamma(\cdot): \mathbb{R}^+ \to \mathbb{R},\ \gamma(\cdot): \tilde{V} \to \gamma(\tilde{V})$: Dimensionless function of dimensionless molar volume used to express the mechanical energy balance. Its superscripts *IG*, *GC* refer to the IGEOS, GCEOS respectively, while $0\sigma$, $\varepsilon\sigma$ refer to the relevant GCEOS respectively.

$\gamma_k(\cdot): \mathbb{R}^+ \to \mathbb{R},\ \gamma_k(\cdot): \tilde{V} \to \gamma_k(\tilde{V}),\ k = 1,2,3$: Dimensionless components of dimensionless function of dimensionless molar volume used to express the mechanical energy balance.

$\Delta H^\circ_{f,i} \left( \dfrac{J}{mol\ i} \right)$: ith species' standard enthalpy of formation at temperature 298.15 K

$\varepsilon(\cdot)$: Model parameter for GCEOS

$\epsilon_i (J)$: Lennard Jones potential related parameter denoting the maximum energy of attraction between two molecules of the ith species

$\kappa(m)$: Constant defining pipeline roughness used in friction factor relations

$\mu(T) \left( \dfrac{kg\ mix}{m \cdot s} \right)$: Gas mixture viscosity that is a function of temperature

$\mu_i (T) \left( \dfrac{kg\ i}{m \cdot s} \right)$: Viscosity of ith species, that is considered to be a function of temperature

$\nu_{ji}$: Stoichiometric coefficient of the ith species in the jth chemical reaction

$\rho \left( \dfrac{kg\ mix}{m^3\ mix} \right)$: Mass density of nonideal gas mixture

$\rho_0 \left( \dfrac{kg\ mix}{m^3\ mix} \right)$: Mass density of nonideal gas mixture at pipeline inlet

$\rho^{IG}_{0,CH_4} \left( \dfrac{kg\ CH_4}{m^3} \right)$: Mass density of baseline $CH_4$ as ideal gas at pipeline inlet

$\rho_{0,CH_4} \left( \dfrac{kg\ CH_4}{m^3} \right)$: Mass density of baseline $CH_4$ gas at pipeline inlet

$\rho_{0,NG} \left( \dfrac{kg\ NG}{m^3} \right)$: Mass density of NG gas at pipeline inlet

$\sigma(\cdot)$: Model parameter for GC EOS

$\sigma_i \left( \overset{\circ}{A} \right)$: Collision diameter of Lennard Jones potential for ith species

$\Phi_{i,j}(\cdot)$: Dimensionless quantities used in the mixture viscosity semiempirical formula

$\Psi(\cdot)$: Model parameter for GC EOS

$\Omega(\cdot)$: Model parameter for GC EOS

$\Omega_{\mu,i}(\cdot)$: Viscosity Collision integral of ith species

$\omega_i(\cdot)$: Acentric factor of ith species in GC EOS

**Subscripts**

$c$ : Critical point (temperature, pressure, compressibility factor model parameter for GC EOS)

$i, j$ : Species

$k$ : Term of dimensionless parameter-independent function $\gamma(\cdot)$

$r$ : Reduced variable (temperature, pressure)

**Introduction**

      The calculation of pressure drops associated with gas transmission in pipelines has been the subject of several research and teaching efforts. Bird, Stewart, and Lightfoot in [BSL07, p. 464-465] presented a dimensional ODE model and its integral form under the ideal gas assumption, which they subsequently applied for pressure drop prediction to the pipeline transmission of pure methane. Cristello et. al. [Cri23] pursued development of a dimensional gas hydraulic model for simulation of a transmission/distribution pipeline, as well as application of the real-time transient model for leak detection in H2 blended pipelines. They found that up to a 30% blend of H2 can be transported without substantial changes to material or compressor station layout, or up to 60% H2 with compressor upgrading. Zhang et. al. [Zha24] used the Benedict-Webb-Rubin (BWR) equation of state (EOS) in a dimensional model used for power optimization of a single pipeline and pipeline networks with multiple sources. Multi-objective optimization strategies were employed for the multi-source pipeline network. Abbas et. al. [Abb21] employed SRK, PR, Benedict-Webb-Rubin-Starling (BWRS) EOS for thermodynamic calculations. Their resulting dimensional model generated velocity profiles for which the erosional velocity limit of the pipeline was shown not be possible to reach for H2 blends up to 40%. Li, et. al. [Li21] covered the impact of pipeline H2 blending on the Joule-Thomson coefficient of NG, H2 blending ratios from 5% to 30%, employing the SRK, PR, and BWRS EOS. A database of J-T coefficients for 5%-30% H2 blending, 0.5-20MPa pressures, and temperatures of 275, 300, 350K was generated using a dimensional model. Experimental validation of the calculations showed BWRS EOS to be most accurate in identifying the J-T coefficient. Abd et. al. [Abd21] simulated a 94km pipeline ("Ruswil – Griespass string part of the Transitgas project") on Aspen Hysys Version 9 using the Peng-Robinson EOS, assuming effects of compressor stations to be negligible. Notably, they found that viscosity of the mixture increases for up to 2% H2 blend, then decreases as H2 concentration further increases. Elevation effects were studied for changes of 25m uphill and downhill, and found that increase in elevation results in an increase in pressure drop. Dimensional modeling and experimental results from Bainier, et. al. [Bai19] and Allison, et. al. [All21] show that for 85% H2 blends, the pipeline energy flowrate capacity reaches a minimum value and pressure drop achieves a maximum value. Our work first provides an analytical solution to a general, dimensionless, pressure drop prediction model, which only requires pipeline dimensionless inlet pressure and three temperature, geometry and flow related dimensionless parameter information to provide dimensionless and dimensional pressure drop estimates for general nonideal gas mixtures featuring various cubic EOS models, and then illustrates the model's predictions for pure methane-hydrogen and eight species containing natural gas-hydrogen blends.

**<u>Thermodynamic and Transport Mixture Properties</u>**

**Equation of State**

The thermodynamic nature of the considered gas mixtures is captured by the Ideal Gas and Generic Cubic Equations Of State (IGEOS, GCEOS), [SVAS22, p. 77-78, p. 97-98, p. 503],

which determine the gas' pressure $P$ as a function of temperature $T$ and molar volume $V$ as:

IGEOS: $$P: \mathbb{R}^+ \times (0, \infty) \to \mathbb{R}^+, \; P:(T,V) \to P(T,V) \triangleq \frac{RT}{V} \tag{1}$$

GCEOS:
$$\begin{aligned} &P: D_P \to \mathbb{R}, \; P:(T,V) \to P(T,V) \triangleq \frac{RT}{V-b} - \frac{a(T)}{(\sigma-\varepsilon)b} \left[ \frac{1}{(V+\varepsilon b)} - \frac{1}{(V+\sigma b)} \right] \\ &D_P \triangleq \left\{ (T,V) \in \mathbb{R}^+ \times (b, \infty) : V \notin \left\{ \{-\varepsilon b\} \cup \{-\sigma b\} \right\} \cap \mathbb{R}^+ \right\} \\ &b \triangleq \sum_{i=1}^N y_i b_i > 0, \; b_i \triangleq \Omega \frac{RT_{c,i}}{P_{c,i}} > 0, \; T_{r,i} \triangleq \frac{T}{T_{c,i}} > 0, \; \sigma > \varepsilon \\ &a(T) \triangleq \sum_{i=1}^N \sum_{j=1}^N y_i y_j \left( a_i(T) a_j(T) \right)^{1/2} > 0, \; a_i(T) \triangleq \Psi \frac{\alpha(T_{r,i}; \omega_i) R^2 T_{c,i}^2}{P_{c,i}} > 0 \end{aligned} \tag{2}$$

where $R(J/(mol \cdot K)) \triangleq k \cdot N_A = 8.314, k(J/K) \triangleq 1.380649 \cdot 10^{-23}, N_A(1/mol) \triangleq 6.02214076 \cdot 10^{23}$, and $P_{c,i}, T_{c,i}, \omega_i, y_i$ are the ith species' critical pressure, critical temperature, acentric factor, and mole fraction, and $R, k, N_A$ are the Universal Gas Constant, Boltzmann constant, and Avogadro number, [SI19].

The aforementioned IGEOS, GCEOS can also be used to determine the gas' molar volume $V$ as a function of temperature $T$ and pressure $P$ as:

IGEOS: $$V: \mathbb{R}^+ \times (0, \infty) \to \mathbb{R}^+, \; V:(T,P) \to V(T,P) \triangleq \frac{RT}{P} \tag{3}$$

GCEOS:
$$\begin{aligned} &V: \mathbb{R}^+ \times (0, \infty) \to \mathbb{R}^+, V:(T,P) \to V(T,P) : V(T,P) \text{ is largest root of} \\ &V^3 + \left[ (\varepsilon+\sigma-1)b - \frac{RT}{P} \right] V^2 + \left[ (\varepsilon\sigma-\varepsilon-\sigma)b^2 - \frac{RT}{P}(\varepsilon+\sigma)b + \frac{a(T)}{P} \right] V - \left[ \varepsilon\sigma b^3 + \frac{RT}{P} \varepsilon\sigma b^2 + \frac{a(T)}{P} b \right] = 0 \\ &b \triangleq \Omega \sum_{i=1}^N \left( y_i \frac{RT_{c,i}}{P_{c,i}} \right), \; a(T) \triangleq \Psi R^2 \sum_{i=1}^N \sum_{j=1}^N y_i y_j \left( \frac{\alpha\left(\frac{T}{T_{c,i}}; \omega_i\right) T_{c,i}^2}{P_{c,i}} \frac{\alpha\left(\frac{T}{T_{c,j}}; \omega_j\right) T_{c,j}^2}{P_{c,j}} \right)^{1/2} \end{aligned} \tag{4}$$

It is important to point out that the aforementioned IGEOS and GCEOS models do not yield necessarily equal molar volume $V$ values for a gas mixture of a given composition (in mole fraction $\{y_i\}_{i=1}^N$ terms), temperature $T$, and pressure $P$, and do not yield necessarily equal pressure $P$ values for a gas mixture of a given composition (in mole fraction $\{y_i\}_{i=1}^N$ terms), temperature $T$, and molar volume $V$.

The considered GCEOS models are listed in Table 1 below:

Table 1: GCEOS Parameters

| E.O.S. | $\alpha(T_{r,i}; \omega_i)$ | $\sigma$ | $\varepsilon$ | $\Omega$ | $\Psi$ | $Z_c$ |
|---|---|---|---|---|---|---|
| $RK$ (1949) | $(T_{r,i})^{\frac{-1}{2}}$ | 1 | 0 | 0.08664 | 0.42748 | 1/3 |
| $SRK$ (1972) | $\left[1+\left(0.480+1.574\omega_i-0.176\omega_i^2\right)\left(1-(T_{r,i})^{\frac{1}{2}}\right)\right]^2$ | 1 | 0 | 0.08664 | 0.42748 | 1/3 |
| $PR$ (1976) | $\left[1+\left(0.37464+1.54226\omega_i-0.26992\omega_i^2\right)\left(1-(T_{r,i})^{\frac{1}{2}}\right)\right]^2$ | $1+\sqrt{2}$ | $1-\sqrt{2}$ | 0.07780 | 0.45724 | 0.30740 |

The gas mixture's mass density is directly related to the gas mixture's molar mass, and molar volume, which satisfy the following relations:

$$\rho = \frac{M}{V}, \quad M = \sum_{i=1}^{N} M_i y_i, \quad M_i = m_i \cdot N_A, \quad i = 1, N \tag{5}$$

where $\rho, M, V$ are the mixture's mass density, molar mass and molar volume, and $M_i, y_i, m_i$ are the ith species' molar mass, molar fraction, and single molecule's mass.

Introducing the following variables, then yields the dimensionless IGEOS and GCEOS (DIGEOS, DGCEOS) listed below.

$$\begin{aligned} &\beta \triangleq \frac{Pb}{RT}, \quad q \triangleq \frac{a(T)}{bRT}, \quad \tilde{V} \triangleq \frac{V}{b} \\ &b_i^{IG} \triangleq \frac{RT_{c,i}}{P_{c,i}} = \frac{b_i}{\Omega}, \quad b^{IG} \triangleq \sum_{i=1}^{N} y_i b_i^{IG} = \frac{b}{\Omega}, \quad \beta^{IG} \triangleq \frac{Pb^{IG}}{RT} = \frac{\beta}{\Omega}, \quad \tilde{V}^{IG} \triangleq \frac{V}{b^{IG}} = \frac{\Omega V}{b} = \Omega \tilde{V} \end{aligned} \tag{6}$$

DIGEOS: $\quad \beta^{IG} : (0,\infty) \to \mathbb{R}^+, \quad \beta^{IG} : \tilde{V}^{IG} \to \beta^{IG}(\tilde{V}^{IG}) \triangleq \frac{1}{\tilde{V}^{IG}} \tag{7}$

DGCEOS:
$$\begin{aligned} &\beta : D_\beta \to \mathbb{R}^+, \quad \beta : (q,\tilde{V}) \to \beta(q,\tilde{V}) \triangleq \frac{1}{\tilde{V}-1} - \frac{q}{\sigma-\varepsilon}\left[\frac{1}{(\tilde{V}+\varepsilon)} - \frac{1}{(\tilde{V}+\sigma)}\right] \\ &D_\beta \triangleq \left\{(q,\tilde{V}) \in \mathbb{R}^+ \times (1,\infty) : \tilde{V} \notin \{\{-\varepsilon\} \cup \{-\sigma\}\} \cap \mathbb{R}^+\right\} \\ &q \triangleq \frac{a(T)}{bRT} = \frac{\sum_{i=1}^{N}\sum_{j=1}^{N} y_i y_j (a_i(T) a_j(T))^{1/2}}{RT \sum_{i=1}^{N} y_i b_i}, \quad \begin{cases} b_i \triangleq \Omega \frac{RT_{c,i}}{P_{c,i}} > 0, \; T_{r,i} \triangleq \frac{T}{T_{c,i}} > 0, \; \sigma > \varepsilon \\ a_i(T) \triangleq \Psi \frac{\alpha(T_{r,i};\omega_i) R^2 T_{c,i}^2}{P_{c,i}} > 0 \end{cases} \end{aligned} \tag{8}$$

The above DGCEOS establishes the dimensionless nonideal gas mixture pressure $\beta \triangleq \frac{Pb}{RT}$, normalized by temperature, and the GCEOS parameter $b$ accounting for finite molecule size, as a function of the dimensionless nonideal gas mixture's temperature $q$ and dimensionless molar volume $\tilde{V}$, i.e., $\beta(q,\tilde{V})$, which in turn implies that:

$$d\beta = \frac{\partial \beta(q,\tilde{V})}{\partial q}dq + \frac{\partial \beta(q,\tilde{V})}{\partial \tilde{V}}d\tilde{V}, \text{ where}$$

$$\left\{ \begin{array}{l} \dfrac{\partial \beta(q,\tilde{V})}{\partial q} = \dfrac{-1}{\sigma - \varepsilon}\left[ \dfrac{1}{(\tilde{V}+\varepsilon)} - \dfrac{1}{(\tilde{V}+\sigma)} \right] \\ \dfrac{\partial \beta(q,\tilde{V})}{\partial \tilde{V}} = \dfrac{-1}{(\tilde{V}-1)^2} + \dfrac{q}{\sigma - \varepsilon}\left[ \dfrac{1}{(\tilde{V}+\varepsilon)^2} - \dfrac{1}{(\tilde{V}+\sigma)^2} \right] \end{array} \right\}, \sigma > \varepsilon$$

The aforementioned DIGEOS, DGCEOS can also be used to determine the gas' dimensionless molar volume $\tilde{V}$ as a function of dimensionless temperature $q$, and dimensionless pressure $\beta$ as:

DIGEOS: $\boxed{\tilde{V}^{IG}:(0,\infty) \to \mathbb{R}^+, \tilde{V}^{IG}:\beta^{IG} \to \tilde{V}^{IG}\left(\beta^{IG}\right) = \dfrac{1}{\beta^{IG}} = \dfrac{1}{\dfrac{P}{T}\sum_{i=1}^{N}\left(y_i \dfrac{T_{c,i}}{P_{c,i}}\right)}}$  (9)

DGGCEOS:

$$\boxed{\begin{array}{l} \tilde{V}:\mathbb{R}^+ \times (0,\infty) \to \mathbb{R}^+, \tilde{V}:(q,\beta) \to \tilde{V}(q,\beta):\tilde{V}(q,\beta) \text{ is largest root of} \\[4pt] \tilde{V}^3 + \tilde{V}^2\left(-\dfrac{1}{\beta} - 1 + \sigma + \varepsilon\right) + \tilde{V}\left(-\varepsilon - \sigma + \sigma\varepsilon - \dfrac{(\sigma + \varepsilon - q)}{\beta}\right) + \left(-\sigma\varepsilon - \dfrac{\sigma\varepsilon + q}{\beta}\right) = 0 \\[8pt] \beta \triangleq \dfrac{Pb}{RT} = \Omega\sum_{i=1}^{N}\left(y_i \dfrac{P}{P_{c,i}}\dfrac{T_{c,i}}{T}\right), q \triangleq \dfrac{a(T)}{bRT} = \dfrac{\Psi\sum_{i=1}^{N}\sum_{j=1}^{N}y_i y_j \left(\dfrac{\alpha\left(\dfrac{T}{T_{c,i}};\omega_i\right)T_{c,i}^2}{P_{c,i}} \dfrac{\alpha\left(\dfrac{T}{T_{c,j}};\omega_j\right)T_{c,j}^2}{P_{c,j}}\right)^{1/2}}{\Omega T \sum_{i=1}^{N}\left(y_i \dfrac{T_{c,i}}{P_{c,i}}\right)} \end{array}}$$ (10)

Further, it is also important to point out that the aforementioned DIGEOS and DGCEOS models also do not necessarily yield the same dimensionless molar volume $\tilde{V}$ for a gas mixture of a given composition (in mole fraction $\{y_i\}_{i=1}^{N}$ terms) temperature $T$ and pressure $P$. However, given the above definitions of $\beta^{IG}, \tilde{V}^{IG}, b^{IG}, \beta, \tilde{V}, b$, it holds that for a gas mixture of a given composition (in mole fraction $\{y_i\}_{i=1}^{N}$ terms) temperature $T$ and pressure $P$, it holds:

$$\boxed{\beta^{IG} \triangleq \dfrac{Pb^{IG}}{RT} = \sum_{i=1}^{N}\left(y_i \dfrac{P}{P_{c,i}}\dfrac{T_{c,i}}{T}\right), \beta \triangleq \dfrac{Pb}{RT} = \Omega\sum_{i=1}^{N}\left(y_i \dfrac{P}{P_{c,i}}\dfrac{T_{c,i}}{T}\right) = \Omega\dfrac{Pb^{IG}}{RT} = \Omega\beta^{IG}}$$ (11)

while for a gas mixture of a given composition (in mole fraction $\{y_i\}_{i=1}^{N}$ terms) and molar volume $V$ it holds:

$$\boxed{\tilde{V}^{IG} \triangleq \frac{V}{b^{IG}} = \frac{V}{\sum_{i=1}^{N}\left(y_i \frac{RT_{c,i}}{P_{c,i}}\right)}, \quad \tilde{V} \triangleq \frac{V}{b} = \frac{V}{\Omega \sum_{i=1}^{N}\left(y_i \frac{RT_{c,i}}{P_{c,i}}\right)} = \frac{\tilde{V}^{IG}}{\Omega}} \quad (12)$$

It should also be pointed out that, for a given $\beta^{IG} \triangleq \frac{Pb^{IG}}{RT} = \sum_{i=1}^{N}\left(y_i \frac{P}{P_{c,i}} \frac{T_{c,i}}{T}\right)$, the DIGEOS model enables the immediate evaluation of $\tilde{V}^{IG}\left(\beta^{IG}\right) = \frac{1}{\beta^{IG}}$, while the DGCEOS model for the evaluation of $\tilde{V}(q, \beta)$ through solution of the cubic eqn. (10), requires first the evaluation of

$$\beta = \Omega \beta^{IG} = \Omega \frac{Pb^{IG}}{RT} = \Omega \sum_{i=1}^{N}\left(y_i \frac{P}{P_{c,i}} \frac{T_{c,i}}{T}\right)$$ and the evaluation of

$$q \triangleq \frac{a(T)}{bRT} = \frac{\Psi \sum_{i=1}^{N} \sum_{j=1}^{N} y_i y_j \left(\frac{\alpha\left(\frac{T}{T_{c,i}};\omega_i\right) T_{c,i}^2}{P_{c,i}} \frac{\alpha\left(\frac{T}{T_{c,j}};\omega_j\right) T_{c,j}^2}{P_{c,j}}\right)^{1/2}}{\Omega T \sum_{i=1}^{N}\left(y_i \frac{T_{c,i}}{P_{c,i}}\right)}.$$

To facilitate the comparison of molar volume $V$ IGEOS and GCEOS predictions for a gas mixture of a given composition (in mole fraction $\{y_i\}_{i=1}^{N}$ terms), temperature $T$, and pressure $P$, when dimensionless DIGEOS and DGCEOS computations are carried out, the Figures 1, 2 shown below are created, in which the DIGEOS $\tilde{V}^{IG}$, $\beta^{IG}$ values are plotted against the bottom and left axes, while the DGCEOS $\tilde{V}$, $\beta$ values are plotted against the top and right axes, which relate to each other based on the equations $\tilde{V}^{IG} = \Omega \tilde{V}$ and $\beta^{IG} = \frac{\beta}{\Omega}$.

Based on the $\{y_i\}_{i=1}^{N}$, $T$, $P$ information and the DGCEOS model choice, $\beta$, $\beta^{IG}$, $q$ are first calculated. The value of the latter quantity, $q$, identifies the relevant dimensionless DGCEOS iso-$q$ curve, while the $\beta$, $\beta^{IG}$ values lie at the same horizontal line since

$$\beta = \Omega \sum_{i=1}^{N}\left(y_i \frac{P}{P_{c,i}} \frac{T_{c,i}}{T}\right) = \Omega \beta^{IG}.$$ Then, $\tilde{V}$, $\tilde{V}^{IG}$ are identified as the projections of the intersections of the aforementioned horizontal line with the aforementioned DGCEOS iso-$q$ curve and the DIGEOS curve on the top and bottom axes respectively. Then, since $\tilde{V}^{IG} \triangleq \frac{V}{b^{IG}} = \frac{\Omega V}{b} = \Omega \tilde{V}$, if these projections are on the same vertical line, then the IGEOS and GCEOS molar volume predictions would be equal.

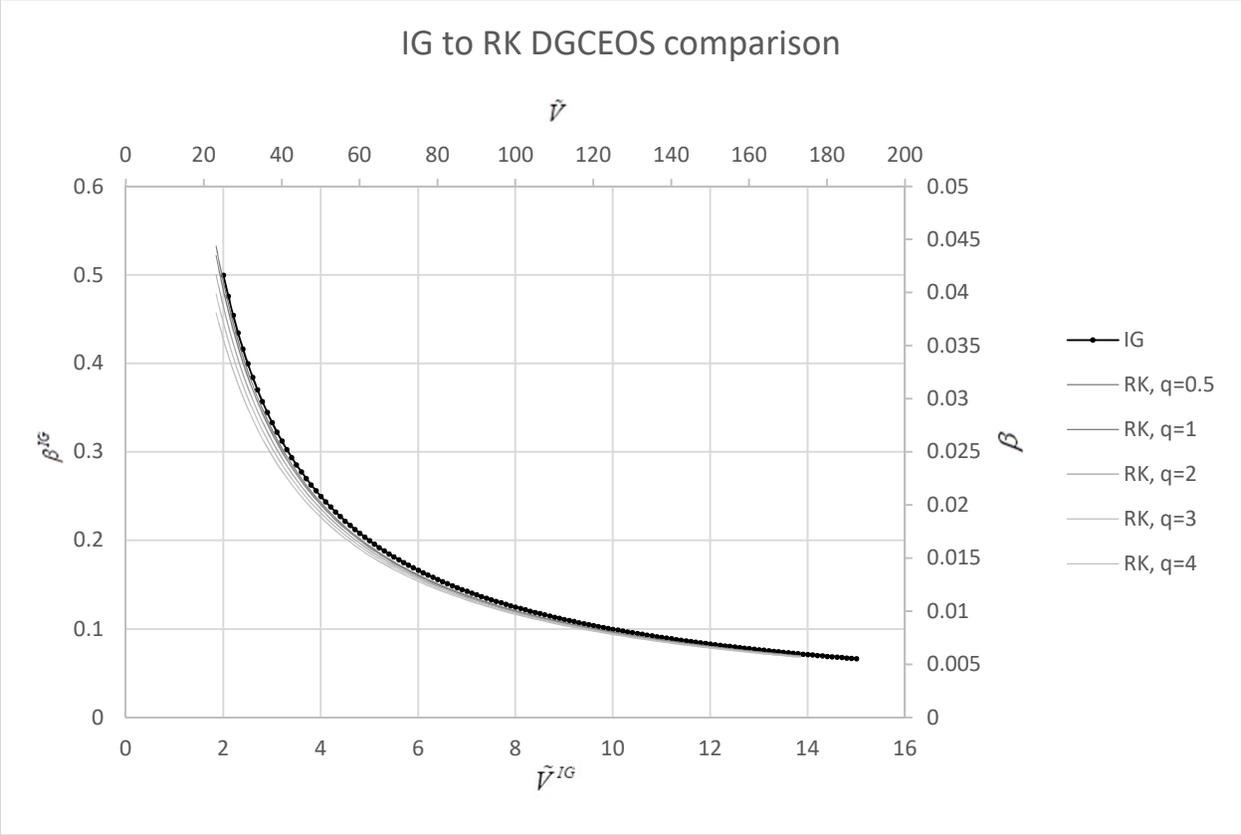

Figure 1: DIGEOS $\tilde{V}^{IG}$, $\beta^{IG}$, Redlich/Kwong, Soave/Redlich/Kwong DGCEOS $\tilde{V}$, $\beta$ iso-$q$ curves

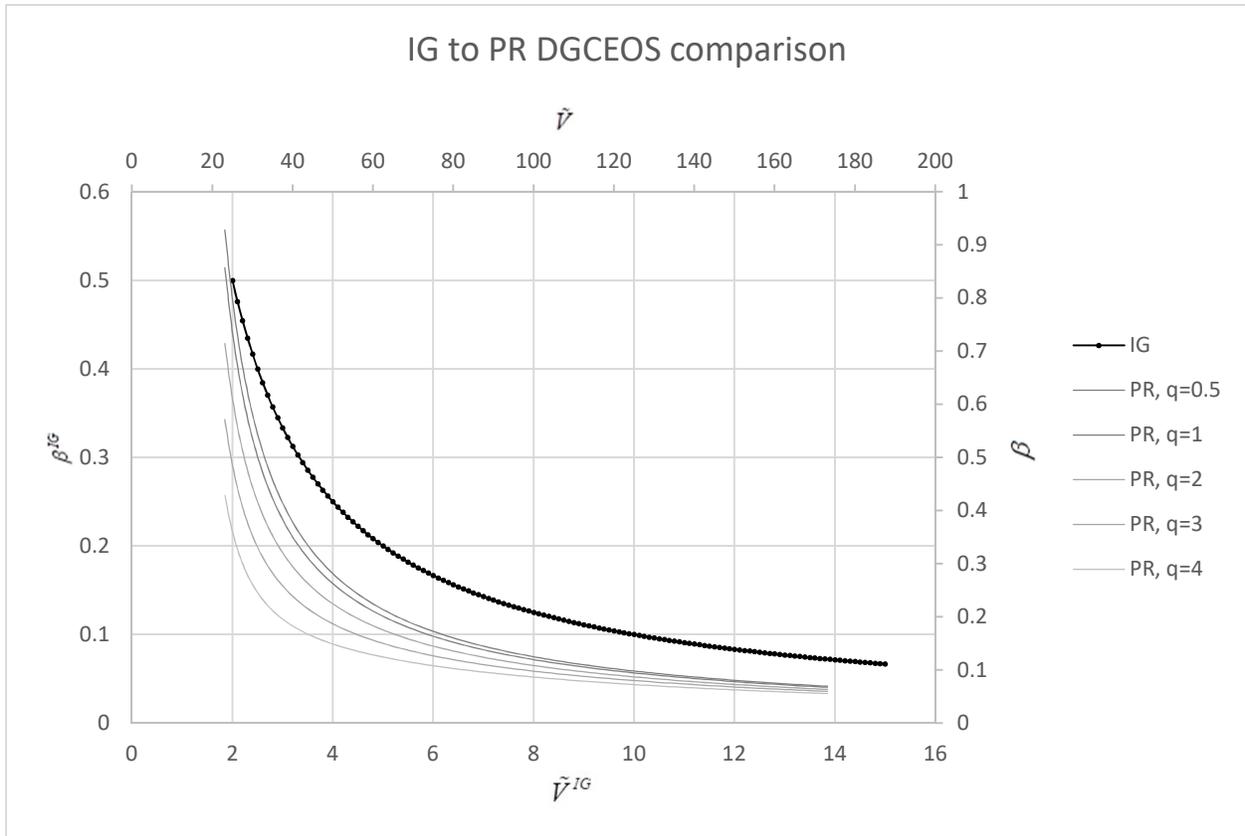

Figure 2: DIGEOS $\tilde{V}^{IG}$, $\beta^{IG}$, Peng/Robinson DGCEOS $\tilde{V}$, $\beta$ iso-$q$ curves

**Viscosity**

The considered nonideal gas mixture is a Newtonian fluid, whose viscosity is considered to be a function of temperature, $\mu(T)$, and is identified by the semiempirical formulas shown below, whose mixing rules were first developed by Wilke [Wil50], modified in [BSL02], and listed in eqns. 1.4-15, 1.4-16, [BSL07, p. 27], and eqns. 12.117, 12.118, [KCG03, p. 518-519], whose species' viscosities $\mu_i(T)$, $i = 1, N$ (also considered to be only functions of temperature) are described by a kinetic theory derived expression listed in eqn. 1.4-14, [BSL07, p. 26], and eqn. 12.100, [KCG03, p. 516], and whose species' viscosity collision integrals $\Omega_{\mu,i}(T)$, $i = 1, N$ are expressed using a curve-fitted expression developed by [NJA72], presented in eqn. E.2-1, [BSL07, p. 866].

$$\boxed{\begin{aligned}
&\mu(T) = \sum_{i=1}^{N} \frac{y_i \mu_i(T)}{\sum_{j=1}^{N} y_j \Phi_{i,j}(T)}, \quad \Phi_{i,j}(T) = \frac{\left[1 + \left(\mu_i(T)/\mu_j(T)\right)^{1/2} \cdot \left(M_j/M_i\right)^{1/4}\right]^2}{\left[8\left(1 + \left(M_i/M_j\right)\right)\right]^{1/2}}, \quad i = 1, N; \; j = 1, N \\
&\mu_i(T) = \frac{5}{16\sqrt{\pi}} \frac{\sqrt{m_i kT}}{\sigma_i^2 \Omega_{\mu,i}(T)} = \frac{5}{16\sqrt{\pi}} \frac{\sqrt{M_i RT}}{N_A \sigma_i^2 \Omega_{\mu,i}(T)} \\
&\Omega_{\mu,i}(T) = \frac{1.16145}{\left(T_i^*\right)^{0.14874}} + \frac{0.52487}{\exp\left(0.77320 \cdot T_i^*\right)} + \frac{2.16178}{\exp\left(2.43787 \cdot T_i^*\right)}, \quad T_i^* \triangleq \frac{kT}{\epsilon_i}
\end{aligned}} \quad (13)$$

where $M_i, y_i, m_i, \sigma_i, T_i^*, \epsilon_i$ are the ith species' molar mass, mole fraction, single molecule's mass, collision diameter of its Lennard Jones potential, dimensionless temperature, and its Lennard Jones potential related parameter denoting the maximum energy of attraction between two molecules of the ith species, and $k$ is the Boltzmann constant. The Lennard Jones potential related parameters $\sigma_i\left(\overset{\circ}{A}\right), \frac{\epsilon_i}{k}(K)$ for various species are provided in Table E.1 of [BSL07, p. 864-865].

**Pipeline model**

This study focuses on the steady state, isothermal flow of a nonideal, compressible, gas mixture flowing through a straight cylindrical pipeline of constant diameter $D_p$, which implies that the pipeline has a constant hydraulic radius $R_h$, and a constant cross section $A_p$, length $L_p$, and differential length $dl$.

Then, using eqn. 6.2-16, [BSL07, p. 183], yields the following relations among the aforementioned pipeline parameters:

$$R_h = \frac{D_p}{4} > 0, \quad A_p = \pi \frac{D_p^2}{4} = 4\pi R_h^2 > 0.$$

**Friction factor model**

The gas mixture's friction factor $f$ can be expressed in terms of the fluid flow's Reynolds number $Re$ defined by eqn. 6.2-18 in [BSL07, p. 183], using the empirical relations by Haaland eqn. 6.2-15 in [BSL07, p. 182], or eqn. 6.21 in [DeN05, p.187]

$$\boxed{\begin{aligned}
&\frac{1}{\sqrt{f}} = -3.6 \cdot \log_{10}\left[\frac{6.9}{Re} + \left(\frac{(\kappa/D_p)}{3.7}\right)^{\frac{10}{9}}\right] \quad if \begin{cases} 4 \cdot 10^4 < Re < 10^8 \\ 0 < \kappa/D_p < 0.05 \end{cases} \\
&f = 0.001375 \cdot \left[1 + \left(20,000(\kappa/D_p) + \frac{10^6}{Re}\right)^{\frac{1}{3}}\right] \\
&Re = \frac{D_p v \rho}{\mu(T)} = \frac{4\dot{m}}{\pi \cdot D_p \cdot \mu(T)}
\end{aligned}} \quad (14)$$

where $\kappa$ is a constant that defines pipeline roughness to be used in friction factor relations.

**Mechanical Energy Balance Model**

Considering that the flow is isothermal, the gas mixture's mechanical energy balance over a pipeline section with no differential height, differential length $dl$ yields the following equation, based on eqn. 15.4-2, [BSL07, p. 461]:

$$vdv + \frac{1}{\rho}dP = d\hat{W} - \frac{1}{2}v^2 \frac{f}{R_h}dl$$

where $v$ is the gas mixture's velocity, $d\hat{W}$ is the differential amount of work per unit mass possibly provided to the gas mixture, and $f$ is the friction factor of the nonideal gas mixture.

Considering no compression equipment exists within the pipeline section yields $d\hat{W} = 0$, and incorporating the aforementioned pipeline geometric relations results in the relation:

$$\rho v dv + dP + \rho \frac{2v^2 f}{D_p} dl = 0.$$

Considering that the gas flow is steady-state, implies that $\dot{m}$ is constant. Given that the cylindrical pipeline has a constant cross-sectional area $A_p$, and denoting $\rho_0, v_0, V_0$ to be the gas mixture's mass density, velocity and molar volume at the pipeline inlet respectively, then

implies: 
$$\begin{cases} \dot{m} = \rho \cdot v \cdot A_p = \frac{M}{V} \cdot v \cdot A_p = \frac{M}{V} \cdot v \cdot \frac{\pi \cdot D_p^2}{4} \\ \dot{m} = \rho_0 \cdot v_0 \cdot A_p = \frac{M}{V_0} \cdot v_0 \cdot A_p = \frac{M}{V_0} \cdot v_0 \cdot \frac{\pi \cdot D_p^2}{4} \end{cases} \Rightarrow \frac{v}{V} = \frac{v_0}{V_0} = \frac{4\dot{m}}{M \cdot \pi \cdot D_p^2}.$$

Then the differential form of the mechanical energy balance becomes:

$$\frac{M}{V}\left(\frac{4\dot{m}}{M \cdot \pi \cdot D_p^2}\right)Vd\left(\frac{4\dot{m}}{M \cdot \pi \cdot D_p^2}V\right) + \frac{M}{V}\left(\frac{2f}{D_p}\left(\frac{4\dot{m}}{M \cdot \pi \cdot D_p^2}\right)^2 V^2\right)dl + dP = 0 \quad \underset{\dot{m}=constant}{\overset{\{y_i\}_{i=1}^N = constant}{\Rightarrow}}$$

$$M\left(\frac{4\dot{m}}{M \cdot \pi \cdot D_p^2}\right)^2 dV + M\left(\frac{2f}{D_p}\left(\frac{4\dot{m}}{M \cdot \pi \cdot D_p^2}\right)^2 V\right)dl + dP = 0$$

Introducing the additional dimensionless variables listed below, allows the expression of the above relation in dimensionless form utilizing the IGEOS, GCEOS, and DIGEOS, DGCEOS.

$$\begin{aligned}
&\tilde{l} \triangleq \frac{l}{L_p} \in (0,1), \tilde{v} \triangleq \frac{v}{v_0} > 0, \ \tilde{V}_0^{IG} \triangleq \frac{V_0}{b^{IG}} = \frac{v_0 \cdot M \cdot \pi \cdot D_p^{\,2}}{4\dot{m}b^{IG}} \\
&\tilde{V}_0 \triangleq \frac{V_0}{b} = \frac{v_0 \cdot M \cdot \pi \cdot D_p^{\,2}}{4\dot{m}b} = \frac{v_0 \cdot M \cdot \pi \cdot D_p^{\,2}}{4\dot{m}\Omega b^{IG}} = \frac{\tilde{V}_0^{IG}}{\Omega} \\
&r_{LD} \triangleq \frac{D_p}{L_p} \in (0,1), \ r_{KI}^{IG} \triangleq \frac{M}{RT}\left(\frac{4\dot{m}b^{IG}}{M \cdot \pi \cdot D_p^{\,2}}\right)^2 = \frac{M}{RT}\left(\frac{v_0}{\tilde{V}_0^{IG}}\right)^2 \\
&r_{KI} \triangleq \frac{M}{RT}\left(\frac{4\dot{m}b}{M \cdot \pi \cdot D_p^{\,2}}\right)^2 = \frac{M}{RT}\left(\frac{v_0}{\tilde{V}_0}\right)^2 = \frac{\Omega^2 M}{RT}\left(\frac{v_0}{\tilde{V}_0^{IG}}\right)^2 = \Omega^2 \, r_{KI}^{IG}
\end{aligned} \quad (15)$$

The Theorem below provides the dimensionless expressions of the integrated form of the mechanical energy balance. Its proof is provided in Appendix A.1.

<u>Theorem</u>

The integrated dimensionless forms of the mechanical energy balance (DIGMEB, DGCMEB) for the isothermal, steady-state flow of a gas mixture over a horizontal, cylindrical pipeline are:

a. DIGMEB for DIGEOS:

$$\begin{aligned}
&\gamma^{IG}\left(\tilde{V}^{IG}(1)\right) - \gamma^{IG}\left(\tilde{V}_0^{IG}\right) = \frac{2f}{r_{LD}}, \ \gamma^{IG}\left(\tilde{V}^{IG}\right) \triangleq \gamma_1\left(\tilde{V}^{IG}\right) + \frac{1}{r_{KI}^{IG}} \cdot \gamma_2^{IG}\left(\tilde{V}^{IG}\right) \ \forall \tilde{V}^{IG} \\
&\left\{\gamma_1\left(\tilde{V}^{IG}\right) \triangleq -ln\left(\tilde{V}^{IG}\right), \ \gamma_2^{IG}\left(\tilde{V}^{IG}\right) \triangleq \frac{-1}{2\left(\tilde{V}^{IG}\right)^2}, \ \gamma^{IG}\left(\tilde{V}^{IG}\right) \triangleq \gamma_1\left(\tilde{V}^{IG}\right) + \frac{1}{r_{KI}^{IG}}\gamma_2^{IG}\left(\tilde{V}^{IG}\right)\right\} \forall \tilde{V}^{IG}
\end{aligned} \quad (16)$$

b. DGCMEB for GCEOS:

$$\boxed{\begin{aligned}
& \gamma^{\varepsilon\sigma}\left(\tilde{V}(1)\right) - \gamma^{\varepsilon\sigma}\left(\tilde{V}_0\right) = \frac{2f}{r_{LD}} \text{ if } \begin{Bmatrix} \varepsilon \neq 0 \\ \sigma \neq 0 \end{Bmatrix} \\
& \gamma^{0\sigma}\left(\tilde{V}(1)\right) - \gamma^{0\sigma}\left(\tilde{V}_0\right) = \frac{2f}{r_{LD}} \text{ if } \begin{Bmatrix} \varepsilon = 0 \\ \sigma \neq 0 \end{Bmatrix} \\
& \gamma^{\varepsilon\sigma}\left(\tilde{V}\right) \triangleq \left[ \gamma_1\left(\tilde{V}\right) + \frac{1}{r_{KI}} \cdot \gamma_2^{GC}\left(\tilde{V}\right) + \frac{q}{r_{KI}} \cdot \gamma_3^{\varepsilon\sigma}\left(\tilde{V}\right) \right] \forall \tilde{V} \text{ if } \begin{Bmatrix} \varepsilon \neq 0 \\ \sigma \neq 0 \end{Bmatrix} \\
& \gamma^{0\sigma}\left(\tilde{V}\right) \triangleq \left[ \gamma_1\left(\tilde{V}\right) + \frac{1}{r_{KI}} \cdot \gamma_2^{GC}\left(\tilde{V}\right) + \frac{q}{r_{KI}} \cdot \gamma_3^{0\sigma}\left(\tilde{V}\right) \right] \forall \tilde{V} \text{ if } \begin{Bmatrix} \varepsilon = 0 \\ \sigma \neq 0 \end{Bmatrix} \\
& \gamma_1\left(\tilde{V}\right) \triangleq -\ln\left(\tilde{V}\right) \forall \tilde{V}, \ \gamma_2^{GC}\left(\tilde{V}\right) \triangleq \ln\left(\frac{\tilde{V}}{\tilde{V}-1}\right) - \frac{1}{\left(\tilde{V}-1\right)} \forall \tilde{V}, \\
& \gamma_3^{\varepsilon\sigma}\left(\tilde{V}\right) \triangleq \frac{-1}{\varepsilon^2(\sigma-\varepsilon)}\left[\ln\left(\frac{\tilde{V}}{\varepsilon+\tilde{V}}\right) + \frac{\varepsilon}{\varepsilon+\tilde{V}}\right] + \frac{1}{\sigma^2(\sigma-\varepsilon)}\left[\ln\left(\frac{\tilde{V}}{\sigma+\tilde{V}}\right) + \frac{\sigma}{\sigma+\tilde{V}}\right] \forall \tilde{V} \text{ if } \begin{Bmatrix} \varepsilon \neq 0 \\ \sigma \neq 0 \end{Bmatrix} \\
& \gamma_3^{0\sigma}\left(\tilde{V}\right) \triangleq \frac{1}{\sigma^2}\left[\frac{1}{\sigma}\ln\left(\frac{\tilde{V}}{\sigma+\tilde{V}}\right) + \frac{1}{\sigma+\tilde{V}}\right] + \frac{1}{2\sigma\left(\tilde{V}\right)^2} \forall \tilde{V} \text{ if } \begin{Bmatrix} \varepsilon = 0 \\ \sigma \neq 0 \end{Bmatrix}
\end{aligned}} \quad (17)$$

The Theorem's application to the considered DIGEOS, DGCEOS models yields the table below:

Table 2: $\gamma^{IG}$, $\gamma^{0\sigma}$, $\gamma^{\varepsilon\sigma}$ functions involved in DIGMEB, DGCMEB models

| | |
|---|---|
| *Ideal Gas* | $\gamma^{IG}\left(\tilde{V}^{IG}\right) \triangleq \left[\gamma_1\left(\tilde{V}^{IG}\right) + \dfrac{1}{r_{KI}^{IG}}\gamma_2^{IG}\left(\tilde{V}^{IG}\right)\right] = \left[-ln\left(\tilde{V}^{IG}\right) + \dfrac{1}{r_{KI}^{IG}}\left(\dfrac{-1}{2\left(\tilde{V}^{IG}\right)^2}\right)\right]$ |
| *RK*(1949) | $\gamma^{0\sigma}\left(\tilde{V}\right) \triangleq \begin{bmatrix}\gamma_1\left(\tilde{V}\right) + \\ +\dfrac{1}{r_{KI}} \cdot \gamma_2^{GC}\left(\tilde{V}\right) + \\ +\dfrac{q}{r_{KI}} \cdot \gamma_3^{0\sigma}\left(\tilde{V}\right)\end{bmatrix} = \left[-ln\left(\tilde{V}\right) + \dfrac{1}{r_{KI}} \cdot \begin{bmatrix}ln\left(\dfrac{\tilde{V}}{\tilde{V}-1}\right) - \\ -\dfrac{1}{\left(\tilde{V}-1\right)}\end{bmatrix} + \dfrac{q}{r_{KI}} \cdot \left[\left[ln\left(\dfrac{\tilde{V}}{\tilde{V}+1}\right) + \dfrac{1}{\tilde{V}+1} + \dfrac{1}{2\left(\tilde{V}\right)^2}\right]\right]\right]$ |
| *SRK*(1972) | $\gamma^{0\sigma}\left(\tilde{V}\right) \triangleq \begin{bmatrix}\gamma_1\left(\tilde{V}\right) + \\ +\dfrac{1}{r_{KI}} \cdot \gamma_2^{GC}\left(\tilde{V}\right) + \\ +\dfrac{q}{r_{KI}} \cdot \gamma_3^{0\sigma}\left(\tilde{V}\right)\end{bmatrix} = \left[-ln\left(\tilde{V}\right) + \dfrac{1}{r_{KI}} \cdot \begin{bmatrix}ln\left(\dfrac{\tilde{V}}{\tilde{V}-1}\right) - \\ -\dfrac{1}{\left(\tilde{V}-1\right)}\end{bmatrix} + \dfrac{q}{r_{KI}} \cdot \left[\left[ln\left(\dfrac{\tilde{V}}{\tilde{V}+1}\right) + \dfrac{1}{\tilde{V}+1} + \dfrac{1}{2\left(\tilde{V}\right)^2}\right]\right]\right]$ |
| *PR*(1976) | $\gamma^{\varepsilon\sigma}\left(\tilde{V}\right) \triangleq \begin{bmatrix}\gamma_1\left(\tilde{V}\right) + \\ +\dfrac{1}{r_{KI}} \cdot \gamma_2^{GC}\left(\tilde{V}\right) + \\ +\dfrac{q}{r_{KI}} \cdot \gamma_3^{\varepsilon\sigma}\left(\tilde{V}\right)\end{bmatrix} = \left[-ln\left(\tilde{V}\right) + \dfrac{1}{r_{KI}} \cdot \begin{bmatrix}ln\left(\dfrac{\tilde{V}}{\tilde{V}-1}\right) - \\ -\dfrac{1}{\left(\tilde{V}-1\right)}\end{bmatrix} + \dfrac{q}{r_{KI}} \cdot \begin{bmatrix}\dfrac{-1}{\left(3-2\sqrt{2}\right)\cdot 2\sqrt{2}}\begin{bmatrix}ln\left(\dfrac{\tilde{V}}{1-\sqrt{2}+\tilde{V}}\right) + \\ +\dfrac{1-\sqrt{2}}{1-\sqrt{2}+\tilde{V}}\end{bmatrix} + \\ +\dfrac{1}{\left(3+2\sqrt{2}\right)\cdot 2\sqrt{2}}\begin{bmatrix}ln\left(\dfrac{\tilde{V}}{1+\sqrt{2}+\tilde{V}}\right) + \\ +\dfrac{1+\sqrt{2}}{1+\sqrt{2}+\tilde{V}}\end{bmatrix}\end{bmatrix}\right]$ |

In advance of the case study computations, it is important to assess the behavior of the $\gamma^{IG}\left(\tilde{V}^{IG}\right)$ function involved in the DIGMEB for DIGEOS. The following holds:

$$\gamma^{IG}\left(\tilde{V}^{IG}\right) \triangleq \gamma_1\left(\tilde{V}^{IG}\right) + \frac{1}{r_{KI}^{IG}}\gamma_2^{IG}\left(\tilde{V}^{IG}\right) = -\ln\left(\tilde{V}^{IG}\right) - \frac{1}{2r_{KI}^{IG}\left(\tilde{V}^{IG}\right)^2} \Rightarrow$$

$$\frac{d\gamma^{IG}\left(\tilde{V}^{IG}\right)}{d\tilde{V}^{IG}} = -\frac{1}{\tilde{V}^{IG}} + \frac{1}{r_{KI}^{IG}\left(\tilde{V}^{IG}\right)^3} \Rightarrow \frac{d\gamma^{IG}\left(\tilde{V}^{IG}\right)}{d\tilde{V}^{IG}} \begin{cases} > 0 \ \forall \tilde{V}^{IG} \in \left(0, \frac{1}{\sqrt{r_{KI}^{IG}}}\right) \\ = 0 \ \forall \tilde{V}^{IG} = \frac{1}{\sqrt{r_{KI}^{IG}}} \\ < 0 \ \forall \tilde{V}^{IG} \in \left(\frac{1}{\sqrt{r_{KI}^{IG}}}, +\infty\right) \end{cases}.$$

This implies that the $\gamma^{IG}\left(\tilde{V}^{IG}\right)$ function is strictly monotonically increasing over $\left(0, \frac{1}{\sqrt{r_{KI}^{IG}}}\right)$, strictly monotonically decreasing over $\left(\frac{1}{\sqrt{r_{KI}^{IG}}}, +\infty\right)$, and has a maximum at $\tilde{V}^{IG} = \frac{1}{\sqrt{r_{KI}^{IG}}}$ whose value is: $\max_{\tilde{V}^{IG} \in (0,+\infty)} \gamma^{IG}\left(\tilde{V}^{IG}\right) = \gamma^{IG}\left(\frac{1}{\sqrt{r_{KI}^{IG}}}\right) = \frac{1}{2}\left(\ln\left(r_{KI}^{IG}\right) - 1\right) \begin{cases} > 0 \ if \ r_{KI}^{IG} \in (e, +\infty) \\ = 0 \ if \ r_{KI}^{IG} = e \\ < 0 \ if \ r_{KI}^{IG} \in (0, e) \end{cases}$.

Further, $\lim_{\tilde{V}^{IG} \to +\infty} \gamma^{IG}\left(\tilde{V}^{IG}\right) = -\infty$, $\lim_{\tilde{V} \to +\infty} \gamma^{0\sigma}\left(\tilde{V}\right) = -\infty$, [Wol24c].

Given that the DIGMEB for DIGEOS has the form $\gamma^{IG}\left(\tilde{V}^{IG}(1)\right) - \gamma^{IG}\left(\tilde{V}_0^{IG}\right) = \frac{2f}{r_{LD}} = \frac{2fL_p}{D_p} > 0$, it is clear that it must hold that $\gamma^{IG}\left(\tilde{V}^{IG}(1)\right) > \gamma^{IG}\left(\tilde{V}_0^{IG}\right)$. At any intermediate pipeline length $l < L_p$, DIGMEB for DIGEOS has the form $\gamma^{IG}\left(\tilde{V}^{IG}\left(\frac{l}{L_p}\right)\right) - \gamma^{IG}\left(\tilde{V}_0^{IG}\right) = \frac{2fl}{D_p} > 0$, which means $\gamma^{IG}\left(\tilde{V}^{IG}\left(\frac{l}{L_p}\right)\right)$ is a linearly increasing function of $l$. Further, since the pipeline pressure (and the dimensionless pipeline pressure $\beta^{IG}$) must be decreasing throughout the pipeline, and it holds that $\beta^{IG}\left(\tilde{V}^{IG}\left(\frac{l}{L_p}\right)\right) = \frac{1}{\tilde{V}^{IG}\left(\frac{l}{L_p}\right)}$, it must then hold that $\tilde{V}^{IG}\left(\frac{l}{L_p}\right)$ is a strictly monotonically increasing function of $l$. It therefore holds,

$\gamma^{IG}\left(\tilde{V}^{IG}\left(\dfrac{l_2}{L_p}\right)\right) > \gamma^{IG}\left(\tilde{V}^{IG}\left(\dfrac{l_1}{L_p}\right)\right) \Leftrightarrow l_2 > l_1 \Leftrightarrow \tilde{V}^{IG}\left(\dfrac{l_2}{L_p}\right) > \tilde{V}^{IG}\left(\dfrac{l_1}{L_p}\right)$, which in turn implies that the $\gamma^{IG}\left(\tilde{V}^{IG}\right)$ has to be a strictly monotonically increasing function of $\tilde{V}^{IG} = \dfrac{1}{\sqrt{r_{KI}^{IG}}}$ which based on the earlier monotonicity analysis implies that the DIGMEB computations are physically meaningful when $\tilde{V}^{IG}(1) \in \left(0, \dfrac{1}{\sqrt{r_{KI}^{IG}}}\right)$. The behavior of the $\gamma^{IG}\left(\tilde{V}^{IG}\right)$, $\gamma^{0\sigma}(\tilde{V})$, $\gamma^{\varepsilon\sigma}(\tilde{V})$ functions involved in the DIGMEB for DIGEOS, and DGCMEB for RK/SRK and PR GCEOS, within domains in which they are monotonically increasing is illustrated below in the Figures 3, 4, 5 for various iso-$r_{KI}^{IG}$, iso-$(q, r_{KI})$, iso-$(q, r_{KI})$ curves respectively.

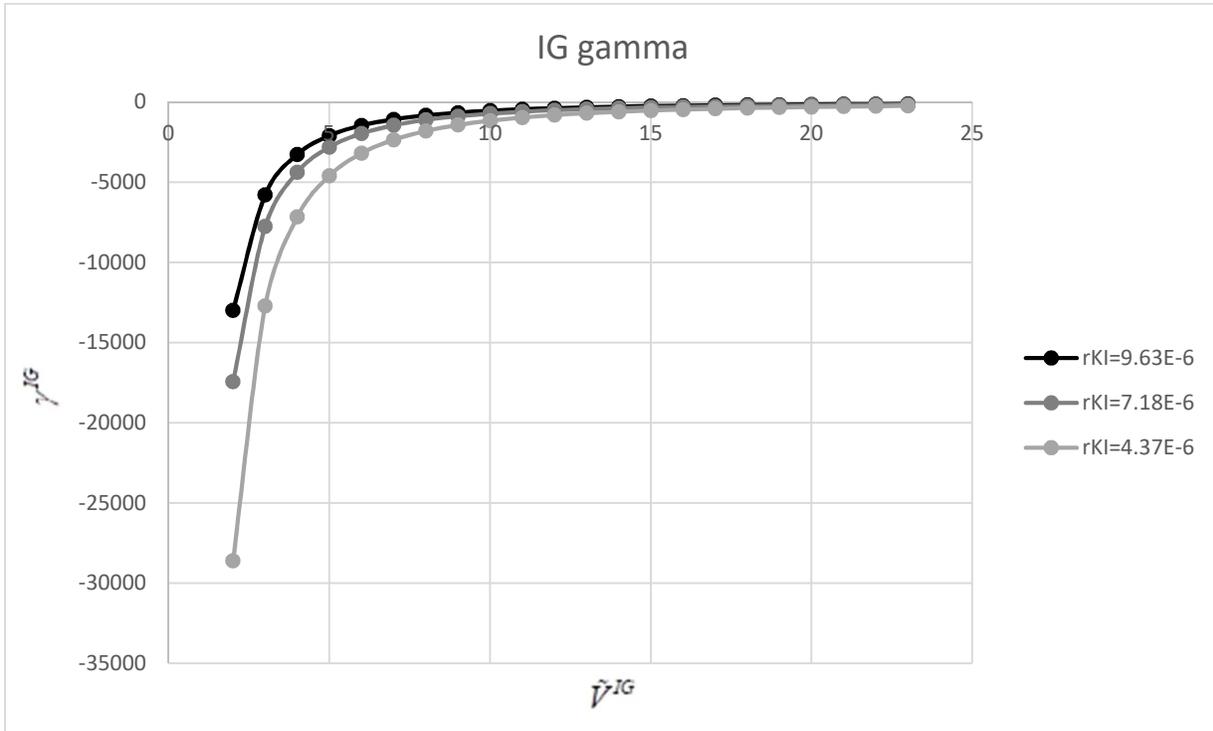

Figure 3: DIGEOS $\tilde{V}^{IG}$, $\gamma^{IG}$ iso-$r_{KI}^{IG}$ curves

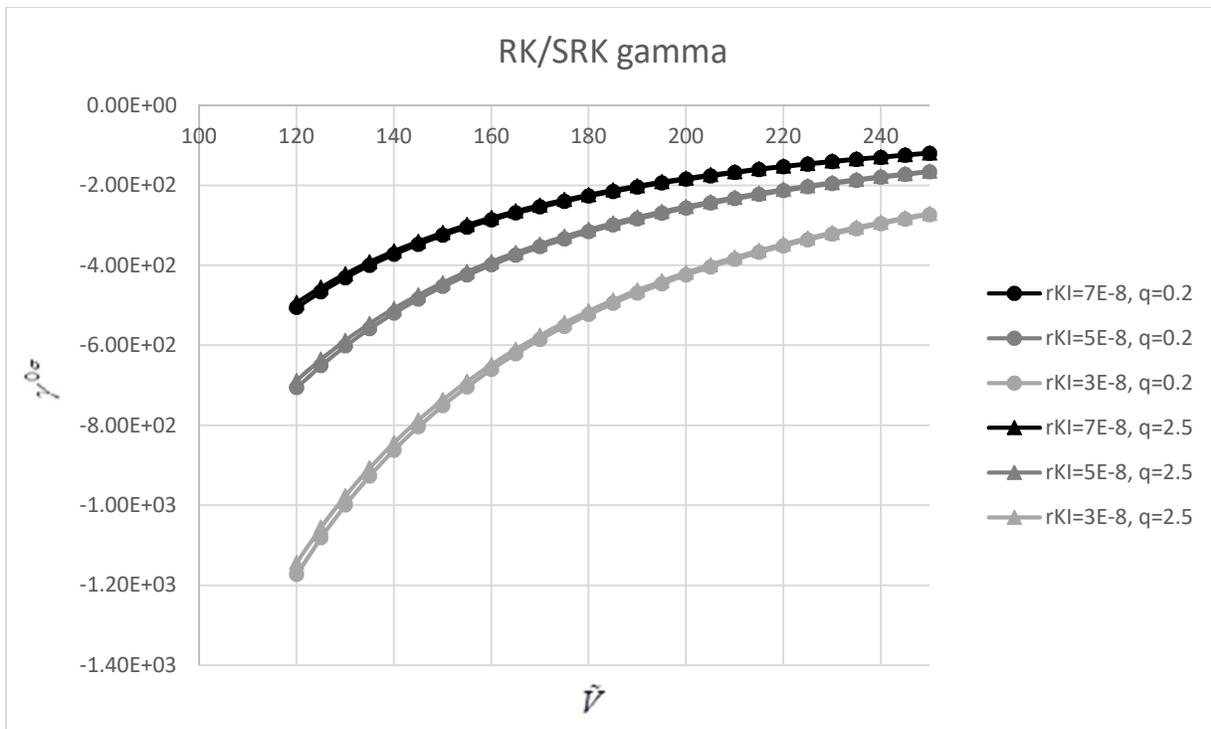

Figure 4: Redlich/Kwong, Soave/Redlich/Kwong DGCEOS $\tilde{V}$, $\gamma^{0\sigma}$ iso-$(q, r_{KI})$ curves

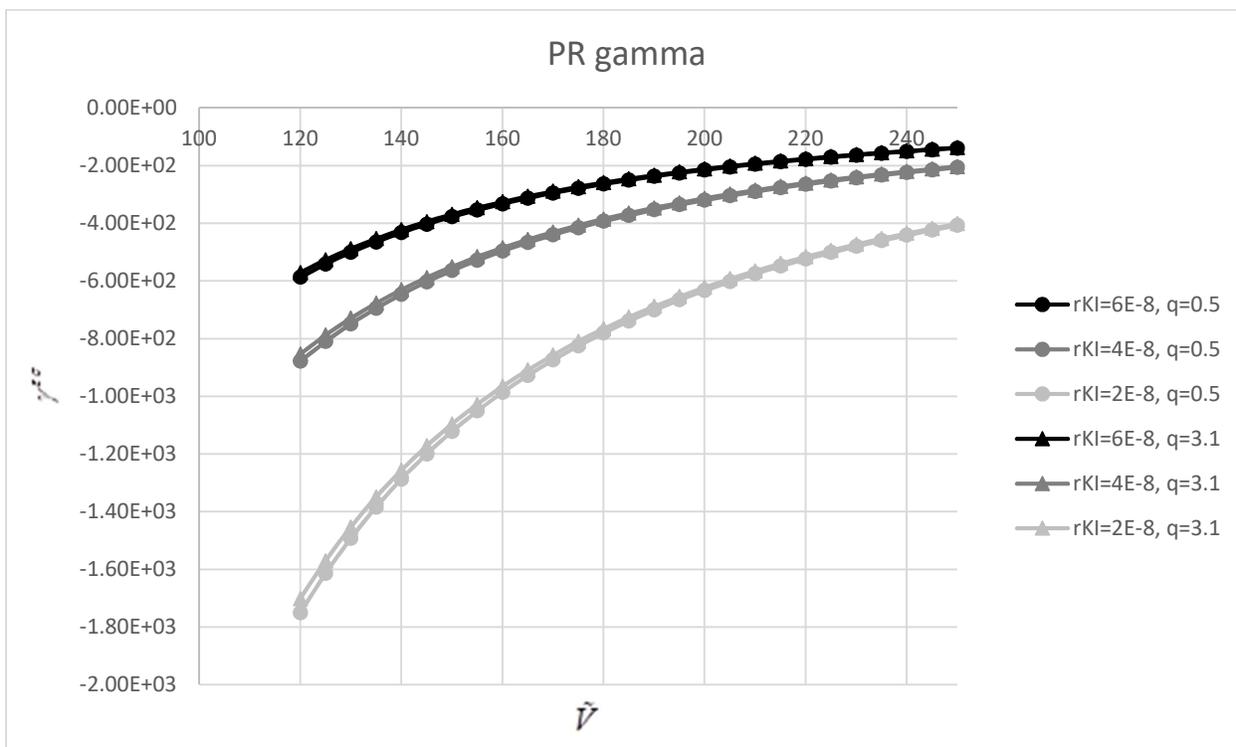

Figure 5: Peng/Robinson DGCEOS $\tilde{V}$, $\gamma^{\varepsilon\sigma}$ iso-$(q, r_{KI})$ curves

**CH4/H2 Blend Pipeline Model and Case Studies**

Two case studies will be conducted for H2 containing mixtures: (1) CH4/H2 and (2) natural gas/H2 (NG/H2), with NG containing the following species: 95.124% CH4, 1.438% N2, 0.530% CO2, 2.721% C2H6, 0.161% C3H8, 0.011% i-C4H10, 0.012% n-C4H10, 0.003% n-C6H14.
In carrying out these two case studies, the species data in Table 3 below will be used. The species' molar masses, acentric factors, critical temperatures and pressures are obtained from [SVAS18, Table B.1, p.650]. The species' standard enthalpies of formation (at temperature 298.15 K (25 °C) and pressure $10^5$ Pa) are obtained from [SVAS18, p. 147, Table C.4, p.658]. The value of the isobutane's standard enthalpy of formation is not included in the aforementioned reference. It is thus obtained from [Wik24] as follows. In SVAS18, p. 147, Table C.4 the enthalpy of formation of $n-C_4H_{10}(g)$ is listed as -125,790 J/mol, while in [Wik24] the enthalpies of formation of $n-C_4H_{10}(g)$ and $i-C_4H_{10}(g)$ are listed as -125.5 J/mol and -134.3 J/mol respectively. Thus, considering the difference of the two enthalpies to be the important metric to be kept constant, the enthalpy of formation of $i-C_4H_{10}(g)$ is listed as -134.590 J/mol.

The species' Lennard Jones potential related parameters $\sigma_i\left(\overset{\circ}{A}\right), \frac{\epsilon_i}{k}(K)$ are obtained from Table E.1 of [BSL07, p. 864-865].

Table 3: Species parameters

| Species i | $M_i$ (kg i/mol i) | $\omega_i$ | $T_{c,i}(K)$ | $P_{c,i}$ (bar) | $\Delta H^\circ_{f,i}$ (J/mol i) | $\sigma_i\left(\overset{\circ}{A}\right)$ | $\frac{\epsilon_i}{k}(K)$ |
|---|---|---|---|---|---|---|---|
| $CH_4(g)$ | $16.043 \cdot 10^{-3}$ | 0.012 | 190.6 | 45.99 | $-74,520$ | 3.780 | 154 |
| $H_2(g)$ | $2.016 \cdot 10^{-3}$ | $-0.216$ | 33.19 | 13.13 | 0 | 2.915 | 38.0 |
| $O_2(g)$ | $31.999 \cdot 10^{-3}$ | 0.022 | 154.6 | 50.43 | 0 | 3.433 | 113 |
| $N_2(g)$ | $28.014 \cdot 10^{-3}$ | 0.038 | 126.2 | 34.00 | 0 | 3.667 | 99.8 |
| $CO_2(g)$ | $44.010 \cdot 10^{-3}$ | 0.224 | 304.2 | 73.83 | $-393,509$ | 3.996 | 190 |
| $C_2H_6(g)$ | $30.070 \cdot 10^{-3}$ | 0.100 | 305.3 | 48.72 | $-83,820$ | 4.388 | 232 |
| $C_3H_8(g)$ | $44.097 \cdot 10^{-3}$ | 0.152 | 369.8 | 42.48 | $-104,680$ | 4.934 | 273 |
| $i-C_4H_{10}(g)$ | $58.123 \cdot 10^{-3}$ | 0.181 | 408.1 | 36.48 | $-134,590$ | 5.393 | 295 |
| $n-C_4H_{10}(g)$ | $58.123 \cdot 10^{-3}$ | 0.200 | 425.1 | 37.96 | $-125,790$ | 5.604 | 304 |
| $n-C_6H_{14}(g)$ | $86.177 \cdot 10^{-3}$ | 0.301 | 507.6 | 30.25 | $-166,920$ | 6.264 | 342 |
| $H_2O(l)$ | $18.015 \cdot 10^{-3}$ | 0.345 | 647.1 | 220.55 | $-285,830$ | – | – |

The case studies are carried out using the following data information and computational order, so that a non-hydrogen containing gas baseline prediction is created and compared to hydrogen containing gas related predictions:

Provided data: $D_p, L_p, \kappa$ (pipeline diameter, length, surface roughness); $\{y_i\}_{i=1}^N, T, \dot{m}, P_0$, (non-hydrogen containing gas species mole fractions, temperature, and mass flowrate throughout the pipeline, and inlet pressure).

Computational Order: The mass flowrates $\dot{m}$ for all considered hydrogen blends are determined so that all the blends have the same power content, on a Joule per second and HHV basis, as the baseline non-hydrogen containing gas for each case study. The molar mass $M$ for all blends is calculated using the blends' $\{y_i\}_{i=1}^N$ and eqn. 5. $b, b^{IG}$ are first computed for the IGEOS and GCEOS of all blends, using the blends' $\{y_i\}_{i=1}^N$ and eqns. 4, 6 respectively. $a(T), q$ are then computed for the GCEOS and DGCEOS of all blends, using the blends' $\{y_i\}_{i=1}^N$ and $T$ and eqns. 4, 6, and $\beta_0^{IG}, \beta_0 = \Omega \beta_0^{IG}$ are computed for the DIGEOS, DGCEOS of all blends, using the blends' $T$, the calculated $b, b^{IG}$, the inlet pressure $P_0$ and eqn. 6. The dimensionless parameters $r_{KI}^{IG}, r_{KI} = \Omega^2 r_{KI}^{IG}$ are then computed using the pipeline $D_p$, the blends' $\{y_i\}_{i=1}^N$ and $T$, the calculated $\dot{m}, M, b, b^{IG}$ and eqn. 15. The inlet dimensionless molar volumes $\tilde{V}_0^{IG}, \tilde{V}_0$ are then calculated using the calculated $\beta_0^{IG}, \beta_0 = \Omega \beta_0^{IG}, q$ and eqns. 9, 10. The $\dfrac{f}{r_{LD}}$ parameter in eqns. 16, 17 is calculated using the pipeline $D_p, L_p$ and eq. 15 for $r_{LD}, \{y_i\}_{i=1}^N, T$ and eqn. 13 for $\mu(T)$, and $D_p, \kappa$ and the calculated $\dot{m}, \mu(T)$ to calculate $Re, f$ using eqn. 14. Then, $\dfrac{f}{r_{LD}}, \tilde{V}_0^{IG}, \tilde{V}_0, q$, $r_{KI}^{IG}, r_{KI} = \Omega^2 r_{KI}^{IG}$ and eqns. 16, 17 are used to calculate $\tilde{V}^{IG}(1), \tilde{V}(1)$. These quantities are then used to calculate $\beta^{IG}(1), \beta(1)$ using eqns 7, 8, and the IGEOS and GCEOS outlet $P$ predictions using eqn. 6.

The above described computational order is partially illustrated in the IGEOS and GCEOS dependent tables below, which in going from the left to the right columns indicate the order in which the computations are executed, and from top to bottom their dependence on the mixture's hydrogen mole fraction.

**CH4/H2 Case Study**

The pressure drop prediction of the mechanical energy model studied in [BSL 15.4-2 example p.464-465] for pipeline transport of unblended pure CH4 $(y_{H_2} = 0)$ (considered as ideal gas) is used as a baseline for this case study. The relevant data are converted to SI units based on [RAP24], and all subsequent formula based variable/parameter evaluations are carried out using MS-Excel.

Pipeline information (diameter, length, roughness):

$$\begin{cases} D_p = 2\,ft = 0.610m \\ L_p = 10\,mi. = 16093.4m \\ \kappa = 1 \cdot 10^{-5}\,m \end{cases} \Rightarrow \begin{cases} A_p = \pi \dfrac{D_p^{\,2}}{4} = \pi \dfrac{(0.610m)^2}{4} = 0.29225m^2 \\ r_{LD} = \dfrac{D_p}{L_p} = \dfrac{0.610m}{16093.4m} = 3.79037 \cdot 10^{-5} \\ \kappa/D_p = 1 \cdot 10^{-5}/0.610 = 1.64 \cdot 10^{-5} < 0.05 \end{cases}$$

Pipeline CH4 gas temperature: $T = 70°F = 294.261K$

Pipeline CH4 gas inlet pressure, mass flowrate: $P_0 = 100\,psia = 689476Pa, \dot{m} = 16.10967\dfrac{kg\,CH_4}{s}$.

The mass flowrate is calculated using an inlet velocity of $v_0 = 40\frac{ft}{s} = 12.192\frac{m}{s}$, and employing the ideal gas assumption of the aforementioned baseline model.

Employing the above data then yields:

GCEOS Parameter accounting for molecule finite size:

$$\begin{cases} b_{H_2}^{IG} = \dfrac{RT_{c,H_2}}{P_{c,H_2}} = \dfrac{8.314\left(\dfrac{J}{mol\cdot K}\right)\cdot 33.19(K)}{1.313\cdot 10^6(Pa)} = 2.1016\cdot 10^{-4}\left(\dfrac{m^3}{mol}\right) \\ b_{CH_4}^{IG} = \dfrac{RT_{c,CH_4}}{P_{c,CH_4}} = \dfrac{8.314\left(\dfrac{J}{mol\cdot K}\right)\cdot 190.6(K)}{4.599\cdot 10^6(Pa)} = 3.4456\cdot 10^{-4}\left(\dfrac{m^3}{mol}\right) \end{cases} \Rightarrow$$

$$b^{IG} = \begin{bmatrix} 2.1016\cdot 10^{-4}\cdot y_{H_2} + \\ +3.4456\cdot 10^{-4}\cdot(1-y_{H_2}) \end{bmatrix}.$$

Pipeline CH$_4$ ideal gas inlet molar volume and dimensionless molar volume:

$$V_0^{IG} = \dfrac{RT}{P_0} = \dfrac{8.314\left(\dfrac{J}{mol\cdot K}\right)\cdot 294.261(K)}{689476(Pa)} = 35.4833\cdot 10^{-4}\left(\dfrac{m^3}{mol}\right),$$

$$\tilde{V}_0^{IG} = \dfrac{RT}{P_0 b^{IG}} = \dfrac{8.314\left(\dfrac{J}{mol\cdot K}\right)\cdot 294.261(K)}{689476(Pa)\cdot 3.4456\cdot 10^{-4}\left(\dfrac{m^3}{mol}\right)} = 10.2981$$

The mass flowrates for blends of CH$_4$/H$_2$ are determined based on the Higher Heating Value (HHV) on a mass basis. The HHV of a fuel is its standard heat of combustion at 25 °C with liquid water as a product.

The H$_2$ and CH$_4$ combustion reactions with liquid water as products are

$$R_1 : CH_4(g) + 2O_2(g) \to CO_2(g) + 2H_2O(l)$$
$$R_2 : H_2(g) + \dfrac{1}{2}O_2(g) \to H_2O(l)$$

The standard heat of a reaction $R_j$ $j=1,M$, denoted as $\sum_{i=1}^{N} v_{ji} S_i = 0$, $j=1,M$ with $\{S_i\}_{i=1}^{N}$, $\{v_{ji}\}_{i=1}^{N}$ being the considered species and their stoichiometric coefficients in the $R_j$ $j=1,M$ reaction respectively, with the latter being positive for products and negative for reactants, is expressed by the following equation:

$\Delta H_j^{\circ} = \sum_{i=1}^{N} v_{ji}\Delta H_{f,i}^{\circ}$, $j=1,M$, where $\{\Delta H_{f,i}^{\circ}\}_{i=1}^{N}$ are the standard enthalpies of formation (at temperature 298.15 K (25 °C) and pressure $10^5$Pa) of the considered species.

The standard enthalpies of formation (at temperature 298.15 K (25 °C) and pressure $10^5$ Pa) [SVAS18, p. 147, Table C.4, p.658] and molar masses [SVAS18, Table B.1, p.650] (for the species involved in the above reactions $R_j \; j = 1, M = 2$ are:

$$R_1 : CH_4(g) + 2O_2(g) \rightarrow CO_2(g) + 2H_2O(l),$$

$$HHV_{molar}\; CH_4 = \Delta H_1^\circ = \begin{bmatrix} (-1) \cdot (-74,520) + 0 \cdot 0 + (-2) \cdot 0 + \\ +1 \cdot (-393,509) + 2 \cdot (-285,830) + 0 \cdot (-241,818) \end{bmatrix} = -890,649 \left( \frac{J}{mol\, CH_4} \right)$$

$$R_2 : H_2(g) + \frac{1}{2}O_2(g) \rightarrow H_2O(l)$$

$$HHV_{molar}\; H_2 = \Delta H_3^\circ = \begin{bmatrix} 0 \cdot (-74,520) + (-1) \cdot 0 + \left( \frac{-1}{2} \right) \cdot 0 + \\ +0 \cdot (-393,509) + 1 \cdot (-285,830) + 0 \cdot (-241,818) \end{bmatrix} = -285,830 \left( \frac{J}{mol\, CH_4} \right)$$

Considering a H₂, CH₄ blend with the H₂, CH₄ mole fractions denoted as $y_{H_2}$, $y_{CH_4} = 1 - y_{H_2}$ then yields the following formula for the blend's HHV value on a molar basis:

$$R_3 : (1 - y_{H_2}) CH_4(g) + y_{H_2} H_2(g) + \left( 2 - \frac{3}{2} y_{H_2} \right) O_2(g) \rightarrow (1 - y_{H_2}) CO_2(g) + (2 - y_{H_2}) H_2O(l),$$

$$HHV_{molar}\; blend = \Delta H_5^\circ = \begin{bmatrix} -(1 - y_{H_2}) \cdot (-74,520) - y_{H_2} \cdot 0 - \left( 2 - \frac{3}{2} y_{H_2} \right) \cdot 0 + \\ +(1 - y_{H_2}) \cdot (-393,509) + (2 - y_{H_2}) \cdot (-285,830) + 0 \cdot (-241,818) \end{bmatrix} =$$

$$= \left[ 74,520 - 393,509 - 2 \cdot 285,830 + (285,830 + 393,509 - 74,520) y_{H_2} \right] \Rightarrow$$

$$\boxed{HHV_{molar}\; blend = (-890,649 + 604,819 y_{H_2}) \left( \frac{J}{mol\, blend} \right)}$$

HHV values can be readily transformed from a molar basis to a mass basis through division with the average molar mass $M_{blend} \left( \frac{kg\, blend}{mol\, blend} \right)$ of the blend which is equal to:

$$M_{blend} \left( \frac{kg\, blend}{mol\, blend} \right) = y_{CH_4} \left( \frac{mol\, CH_4}{mol\, blend} \right) M_{CH_4} \left( \frac{kg\, CH_4}{mol\, CH_4} \right) + y_{H_2} \left( \frac{mol\, H_2}{mol\, blend} \right) M_{H_2} \left( \frac{kg\, H_2}{mol\, H_2} \right) =$$

$$= (1 - x_{H_2}) 16.043 \cdot 10^{-3} + y_{H_2} 2.016 \cdot 10^{-3} \Rightarrow$$

$$\boxed{M_{blend} \left( \frac{kg\, blend}{mol\, blend} \right) = \left( 16.043 \cdot 10^{-3} - 14.027 \cdot 10^{-3} y_{H_2} \right)}$$

Then, the blend's HHV value on a mass basis are:

$$HHV_{mass}\ blend\left(\frac{J}{kg\ blend}\right) = \frac{HHV_{molar}\ blend\left(\frac{J}{mol\ blend}\right)}{M_{blend}\left(\frac{kg\ blend}{mol\ blend}\right)} \Rightarrow$$

$$\boxed{HHV_{mass}\ blend\left(\frac{J}{kg\ blend}\right) = \frac{(-890,649 + 604,819\,y_{H_2})}{(16.043\cdot 10^{-3} - 14.027\cdot 10^{-3}\,y_{H_2})}}$$

Then the power content of the pure CH$_4$ pipeline inlet, on a Joule per second and HHV basis, is:

$$\dot{m}\cdot HHV_{mass}\ blend = 16.10967\left(\frac{kg\ CH_4}{s}\right)\cdot\frac{(-890,649)}{(16.043\cdot 10^{-3})}\left(\frac{J}{kg\ CH_4}\right) = -8.94350011\cdot 10^8\,\frac{J}{s}.$$

Then, the CH$_4$/H$_2$ blend mass flowrate $\dot{m}(y_{H_2})$ is determined as a function of the hydrogen mole fraction $y_{H_2} \neq 0$, so that the CH$_4$/H$_2$ blend's power content, on a Joule per second and HHV basis, is equal to that of the pure CH$_4$ pipeline inlet. Thus,

$$\dot{m}(y_{H_2})\cdot\frac{(-890,649 + 604,819\,y_{H_2})}{(16.043\cdot 10^{-3} - 14.027\cdot 10^{-3}\,y_{H_2})} = \left(-8.94350011\cdot 10^8\,\frac{J}{s}\right) \Rightarrow$$

$$\boxed{\dot{m}(y_{H_2}) = (-8.94350011\cdot 10^5)\frac{(16.043 - 14.027\,y_{H_2})}{(-890,649 + 604,819\,y_{H_2})}}$$

The species Lennard-Jones parameters provided in the above table, and the species viscosity relations provided earlier, yield the following relations for CH$_4$, H$_2$:

$$\left\{\begin{array}{l}\Omega_{\mu,H_2}(T) = \left[\dfrac{1.16145}{\left(\left(\dfrac{k}{\epsilon_i}\right)_{H_2} T\right)^{0.14874}} + \\ + \dfrac{0.52487}{\exp\left(0.77320 \cdot \left(\dfrac{k}{\epsilon_i}\right)_{H_2} T\right)} + \\ + \dfrac{2.16178}{\exp\left(2.43787 \cdot \left(\dfrac{k}{\epsilon_i}\right)_{H_2} T\right)}\right] = \left[\dfrac{1.16145}{\left(\dfrac{1}{38.0(K)} \cdot 294.261(K)\right)^{0.14874}} + \\ + \dfrac{0.52487}{\exp\left(\dfrac{0.77320}{38.0(K)} \cdot 294.261(K)\right)} + \\ + \dfrac{2.16178}{\exp\left(\dfrac{2.43787}{38.0(K)} \cdot 294.261(K)\right)}\right] = 0.857918 \\[2em] \Omega_{\mu,CH_4}(T) = \left[\dfrac{1.16145}{\left(\left(\dfrac{k}{\epsilon_i}\right)_{CH_4} T\right)^{0.14874}} + \\ + \dfrac{0.52487}{\exp\left(0.77320 \cdot \left(\dfrac{k}{\epsilon_i}\right)_{CH_4} T\right)} + \\ + \dfrac{2.16178}{\exp\left(2.43787 \cdot \left(\dfrac{k}{\epsilon_i}\right)_{CH_4} T\right)}\right] = \left[\dfrac{1.16145}{\left(\dfrac{1}{154(K)} \cdot 294.261(K)\right)^{0.14874}} + \\ + \dfrac{0.52487}{\exp\left(\dfrac{0.77320}{154(K)} \cdot 294.261(K)\right)} + \\ + \dfrac{2.16178}{\exp\left(\dfrac{2.43787}{154(K)} \cdot 294.261(K)\right)}\right] = 1.195096\end{array}\right\}$$

$$\left\{ \begin{array}{l} \mu_{H_2}(T) = \dfrac{5}{16\sqrt{\pi}} \dfrac{\sqrt{M_{H_2}RT}}{N_A \sigma_{H_2}^{\,2} \Omega_{\mu,H_2}(T)} = \dfrac{5}{16\sqrt{\pi}} \sqrt{\dfrac{\left[ 2.016\cdot 10^{-3}\left(\dfrac{kg\,H_2}{mol\,H_2}\right)\cdot 8.314\left(\dfrac{J}{mol\,H_2\cdot K}\right)\cdot 294.261(K) \right]}{\left[ 6.022\cdot 10^{23}\left(\dfrac{1}{mol\,H_2}\right)\cdot \left(2.915\cdot 10^{-10}(m)\right)^2 \cdot 0.857918 \right]}} = 8.9190\cdot 10^{-6}\left(\dfrac{kg\,H_2}{m\cdot s}\right) \\[2em] \mu_{CH_4}(T) = \dfrac{5}{16\sqrt{\pi}} \dfrac{\sqrt{M_{CH_4}RT}}{N_A \sigma_{CH_4}^{\,2} \Omega_{\mu,CH_4}(T)} = \dfrac{5}{16\sqrt{\pi}} \sqrt{\dfrac{\left[ 16.043\cdot 10^{-3}\left(\dfrac{kg\,CH_4}{mol\,CH_4}\right)\cdot 8.314\left(\dfrac{J}{mol\,CH_4\cdot K}\right)\cdot 294.261(K) \right]}{\left[ 6.022\cdot 10^{23}\left(\dfrac{1}{mol\,CH_4}\right)\cdot \left(3.780\cdot 10^{-10}(m)\right)^2 \cdot 1.195096 \right]}} = 1.0741\cdot 10^{-5}\left(\dfrac{kg\,CH_4}{m\cdot s}\right) \end{array} \right.$$

Then from the mixing relations provided earlier, the $CH_4/H_2$ blend viscosity is:

$$\mu(T) = \dfrac{y_{H_2}\mu_{H_2}(T)}{y_{H_2}\Phi_{H_2,H_2}(T) + y_{CH_4}\Phi_{H_2,CH_4}(T)} + \dfrac{y_{CH_4}\mu_{CH_4}(T)}{y_{H_2}\Phi_{CH_4,H_2}(T) + y_{CH_4}\Phi_{CH_4,CH_4}(T)}$$

where:

$$\begin{cases} \Phi_{H_2,H_2}(T) = \dfrac{\left[1+\left(\mu_{H_2}(T)/\mu_{H_2}(T)\right)^{1/2}\cdot\left(M_{H_2}/M_{H_2}\right)^{1/4}\right]^2}{\left[8\left(1+\left(M_{H_2}/M_{H_2}\right)\right)\right]^{1/2}} \\[6pt] \Phi_{CH_4,CH_4}(T) = \dfrac{\left[1+\left(\mu_{CH_4}(T)/\mu_{CH_4}(T)\right)^{1/2}\cdot\left(M_{CH_4}/M_{CH_4}\right)^{1/4}\right]^2}{\left[8\left(1+\left(M_{CH_4}/M_{CH_4}\right)\right)\right]^{1/2}} \\[6pt] \Phi_{H_2,CH_4}(T) = \dfrac{\left[1+\left(\mu_{H_2}(T)/\mu_{CH_4}(T)\right)^{1/2}\cdot\left(M_{CH_4}/M_{H_2}\right)^{1/4}\right]^2}{\left[8\left(1+\left(M_{H_2}/M_{CH_4}\right)\right)\right]^{1/2}} \\[6pt] \Phi_{CH_4,H_2}(T) = \dfrac{\left[1+\left(\mu_{CH_4}(T)/\mu_{H_2}(T)\right)^{1/2}\cdot\left(M_{H_2}/M_{CH_4}\right)^{1/4}\right]^2}{\left[8\left(1+\left(M_{CH_4}/M_{H_2}\right)\right)\right]^{1/2}} \end{cases} \Rightarrow$$

$$\begin{cases} \Phi_{H_2,H_2}(T) = \dfrac{[1+1]^2}{[8(1+1)]^{1/2}}=1,\ \Phi_{CH_4,CH_4}(T) = \dfrac{[1+1]^2}{[8(1+1)]^{1/2}}=1 \\[6pt] \Phi_{H_2,CH_4}(T) = \dfrac{\left[1+\left(8.9190\cdot10^{-6}/1.0741\cdot10^{-5}\right)^{1/2}\cdot\left(16.043\cdot10^{-3}/2.016\cdot10^{-3}\right)^{1/4}\right]^2}{\left[8\left(1+\left(2.016\cdot10^{-3}/16.043\cdot10^{-3}\right)\right)\right]^{1/2}} = 2.1338 \\[6pt] \Phi_{CH_4,H_2}(T) = \dfrac{\left[1+\left(1.0741\cdot10^{-5}/8.9190\cdot10^{-6}\right)^{1/2}\cdot\left(2.016\cdot10^{-3}/16.043\cdot10^{-3}\right)^{1/4}\right]^2}{\left[8\left(1+\left(16.043\cdot10^{-3}/2.016\cdot10^{-3}\right)\right)\right]^{1/2}} = 0.32292 \end{cases}.$$

Thus,

$$\mu(T)(y_{H_2}) = \left[\dfrac{y_{H_2}\cdot 8.9190\cdot 10^{-6}}{y_{H_2}+(1-y_{H_2})\cdot 2.1338} + \dfrac{(1-y_{H_2})\cdot 1.0741\cdot 10^{-5}}{y_{H_2}\cdot 0.32292+(1-y_{H_2})}\right]\left(\dfrac{kg\ blend}{m\cdot s}\right)$$

The above relations $\left(y_{H_2}, \dot{m}(y_{H_2})\right)$, $\left(y_{H_2}, \mu(T)(y_{H_2})\right)$ are illustrated in Figure 6 below. The flowrate $\dot{m}(y_{H_2})$ that delivers the same amount of HHV energy, is a monotonically decreasing function of $y_{H_2}$, while as shown in Appendix A2, the mixture viscosity exhibits a maximum at $y_{H_2} = 0.4613$ and is a monotonically increasing and decreasing function of $y_{H_2}$ before and after $y_{H_2} = 0.4613$. This behavior of mixture viscosity is consistent with findings from the Aspen Hysys pipeline simulations of Abd et. al. [Abd21].

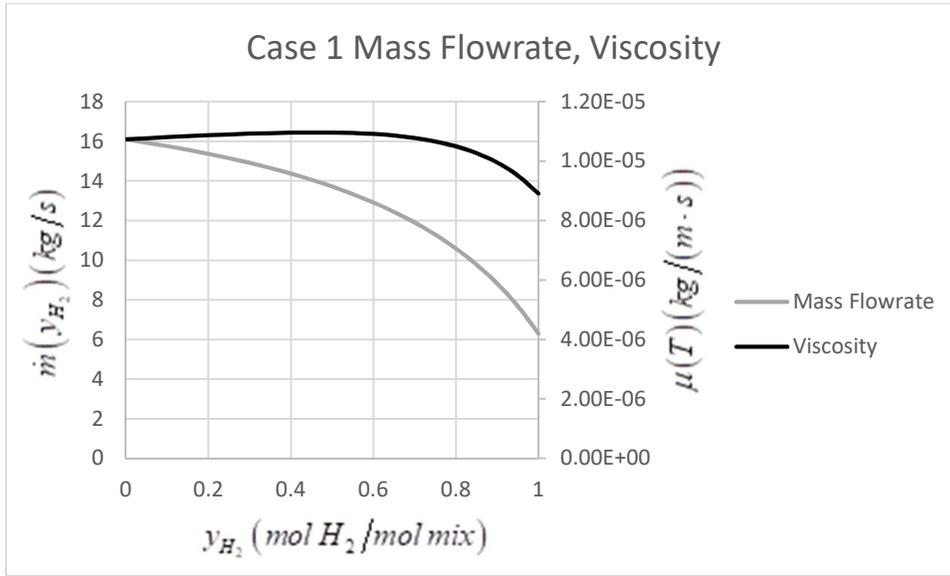

Figure 6: $(y_{H_2}, \dot{m}(y_{H_2}))$, $(y_{H_2}, \mu(T)(y_{H_2}))$ relations for Case 1 CH$_4$/H$_2$ mixture

For the baseline case, the pipeline CH$_4$ gas Reynolds number is:

$$Re = \frac{4\dot{m}}{\pi \cdot D_p \cdot \mu(T)} = \frac{4\dot{m}}{\pi \cdot D_p \cdot \mu_{CH_4}(T)} = \frac{4 \cdot 16.10967 \left(\frac{kg\, CH_4}{s}\right)}{\pi \cdot 0.610(m) \cdot 1.0741 \cdot 10^{-5} \left(\frac{kg\, CH_4}{m \cdot s}\right)} = 3.1305 \cdot 10^6$$

Pipeline CH$_4$ friction factor:

Since $4 \cdot 10^4 < Re < 10^8$ and, considering pipe roughness $\kappa = 1 \cdot 10^{-5}$,
$\kappa/D_p = 1 \cdot 10^{-5}/0.610 = 1.64 \cdot 10^{-5} < 0.05$, the Haaland friction factor model can be used.

$$\frac{1}{\sqrt{f}} = -3.6 \cdot log_{10}\left[\frac{6.9}{Re} + \left(\frac{(\kappa/D_p)}{3.7}\right)^{\frac{10}{9}}\right] \Rightarrow$$

$$f = \left[-3.6 \cdot log_{10}\left[\frac{6.9}{Re} + \left(\frac{(\kappa/D_p)}{3.7}\right)^{\frac{10}{9}}\right]\right]^{-2} = \left[-3.6 \cdot log_{10}\left[\frac{6.9}{3.1305 \cdot 10^6} + \left(\frac{(1.64 \cdot 10^{-5})}{3.7}\right)^{\frac{10}{9}}\right]\right]^{-2} = 2.5717 \cdot 10^{-3}$$

This friction factor value is consistent with the value $f = 2.5 \cdot 10^{-3}$ evaluated in [BSL 15.4-2 example p.464-465]. Then,

$$\frac{2f}{r_{LD}} = \frac{2f}{D_p/L_p} = \frac{2 \cdot 2.5717 \cdot 10^{-3}}{0.610/16093.4} = 1.3570 \cdot 10^2.$$

The resulting pressure drop model predictions for the considered EOS models and for various hydrogen mole fraction $y_{H_2}$ values are:

IG:

| y_H2 | b (IG) | V~_0 | rKI (IG) | V~(1) | P(1) | ΔP |
|---|---|---|---|---|---|---|
| 0 | 0.000345 | 10.29803 | 9.19E-06 | 12.010598 | 591164.82 | -98311.18 |
| 0.05 | 0.000338 | 10.50287 | 9.055E-06 | 12.307656 | 588371.61 | -101104.4 |
| 0.1 | 0.000331 | 10.71602 | 8.916E-06 | 12.620357 | 585438.33 | -104037.7 |
| 0.15 | 0.000324 | 10.93801 | 8.776E-06 | 12.949927 | 582358.12 | -107117.9 |
| 0.2 | 0.000318 | 11.16939 | 8.633E-06 | 13.297694 | 579124.78 | -110351.2 |
| 0.25 | 0.000311 | 11.41077 | 8.487E-06 | 13.665092 | 575733.29 | -113742.7 |
| 0.3 | 0.000304 | 11.66281 | 8.338E-06 | 14.053646 | 572180.68 | -117295.3 |
| 0.35 | 0.000298 | 11.92623 | 8.184E-06 | 14.464954 | 568467.17 | -121008.8 |
| 0.4 | 0.000291 | 12.20183 | 8.025E-06 | 14.900642 | 564597.97 | -124878 |
| 0.45 | 0.000284 | 12.49048 | 7.859E-06 | 15.36229 | 560585.9 | -128890.1 |
| 0.5 | 0.000277 | 12.7931 | 7.685E-06 | 15.851295 | 556455.35 | -133020.6 |
| 0.55 | 0.000271 | 13.11076 | 7.5E-06 | 16.368645 | 552248.17 | -137227.8 |
| 0.6 | 0.000264 | 13.44459 | 7.301E-06 | 16.914549 | 548032.6 | -141443.4 |
| 0.65 | 0.000257 | 13.79587 | 7.085E-06 | 17.487816 | 543917.06 | -145558.9 |
| 0.7 | 0.00025 | 14.166 | 6.846E-06 | 18.08486 | 540071.34 | -149404.7 |
| 0.75 | 0.000244 | 14.55653 | 6.577E-06 | 18.698077 | 536759.95 | -152716.1 |
| 0.8 | 0.000237 | 14.96921 | 6.267E-06 | 19.313268 | 534394.77 | -155081.2 |
| 0.85 | 0.00023 | 15.40597 | 5.901E-06 | 19.905648 | 533619.62 | -155856.4 |
| 0.9 | 0.000224 | 15.86898 | 5.456E-06 | 20.433917 | 535447 | -154029 |
| 0.95 | 0.000217 | 16.36068 | 4.898E-06 | 20.832269 | 541481.96 | -147994 |
| 1 | 0.00021 | 16.88383 | 4.172E-06 | 21.001429 | 554295.48 | -135180.5 |

RK:

| y_H2 | b (RK) | V~_0 | rKI (RK) | a(T) | q | V~(1) | P(1) | ΔP |
|---|---|---|---|---|---|---|---|---|
| 0 | 2.99E-05 | 117.2977 | 7.085E-08 | 0.1878509 | 2.5720692 | 137.38579 | 589790.3 | -99685.7 |
| 0.05 | 2.93E-05 | 119.8161 | 6.979E-08 | 0.1733133 | 2.4202209 | 141.02749 | 586792.2 | -102684 |
| 0.1 | 2.87E-05 | 122.4278 | 6.872E-08 | 0.159361 | 2.2705499 | 144.85186 | 583649.6 | -105826 |
| 0.15 | 2.81E-05 | 125.1386 | 6.764E-08 | 0.1459942 | 2.1231915 | 148.8735 | 580355.5 | -109121 |
| 0.2 | 2.75E-05 | 127.955 | 6.654E-08 | 0.1332128 | 1.9782926 | 153.10824 | 576903.3 | -112573 |
| 0.25 | 2.69E-05 | 130.8837 | 6.541E-08 | 0.1210168 | 1.8360124 | 157.57313 | 573287.9 | -116188 |
| 0.3 | 2.64E-05 | 133.9324 | 6.426E-08 | 0.1094062 | 1.6965246 | 162.28633 | 569506.2 | -119970 |
| 0.35 | 2.58E-05 | 137.1092 | 6.308E-08 | 0.098381 | 1.5600185 | 167.26689 | 565558.3 | -123918 |
| 0.4 | 2.52E-05 | 140.4231 | 6.185E-08 | 0.0879412 | 1.4267006 | 172.53414 | 561449.7 | -128026 |
| 0.45 | 2.46E-05 | 143.8838 | 6.057E-08 | 0.0780868 | 1.2967972 | 178.10683 | 557193.8 | -132282 |
| 0.5 | 2.4E-05 | 147.5022 | 5.923E-08 | 0.0688179 | 1.1705566 | 184.00141 | 552816 | -136660 |
| 0.55 | 2.34E-05 | 151.2898 | 5.781E-08 | 0.0601343 | 1.0482515 | 190.2293 | 548359.8 | -141116 |
| 0.6 | 2.29E-05 | 155.2599 | 5.628E-08 | 0.0520362 | 0.9301827 | 196.7922 | 543896.4 | -145580 |
| 0.65 | 2.23E-05 | 159.4266 | 5.461E-08 | 0.0445234 | 0.8166821 | 203.67458 | 539538.8 | -149937 |

| y_H2 | | | | | | | | |
|---|---|---|---|---|---|---|---|---|
| 0.7 | 2.17E-05 | 163.8059 | 5.277E-08 | 0.0375961 | 0.7081175 | 210.83121 | 535464.2 | -154012 |
| 0.75 | 2.11E-05 | 168.4153 | 5.069E-08 | 0.0312542 | 0.604897 | 218.16723 | 531948.4 | -157528 |
| 0.8 | 2.05E-05 | 173.2746 | 4.83E-08 | 0.0254977 | 0.5074752 | 225.50638 | 529420.9 | -160055 |
| 0.85 | 2E-05 | 178.4055 | 4.548E-08 | 0.0203266 | 0.4163596 | 232.54203 | 528552.6 | -160923 |
| 0.9 | 1.94E-05 | 183.8326 | 4.205E-08 | 0.0157409 | 0.3321189 | 238.76545 | 530397.5 | -159079 |
| 0.95 | 1.88E-05 | 189.5835 | 3.775E-08 | 0.0117406 | 0.2553921 | 243.37106 | 536622.8 | -152853 |
| 1 | 1.82E-05 | 195.6892 | 3.216E-08 | 0.0083257 | 0.1868999 | 245.15567 | 549890.7 | -139585 |

SRK:

| $y_{H2}$ | b (SRK) | V~_0 | rKI (SRK) | a(T) | q | V~(1) | P(1) | ΔP |
|---|---|---|---|---|---|---|---|---|
| 0 | 2.99E-05 | 117.4022 | 7.072E-08 | 0.1803472 | 2.4693281 | 137.46803 | 589883.8 | -99592.2 |
| 0.05 | 2.93E-05 | 119.8985 | 6.967E-08 | 0.1674937 | 2.3389538 | 141.08121 | 586909.8 | -102566 |
| 0.1 | 2.87E-05 | 122.489 | 6.86E-08 | 0.1551153 | 2.2100573 | 144.87749 | 583791.4 | -105685 |
| 0.15 | 2.81E-05 | 125.1795 | 6.752E-08 | 0.143212 | 2.0827303 | 148.87147 | 580521.2 | -108955 |
| 0.2 | 2.75E-05 | 127.9763 | 6.642E-08 | 0.1317839 | 1.9570724 | 153.07897 | 577092.9 | -112383 |
| 0.25 | 2.69E-05 | 130.8865 | 6.53E-08 | 0.1208309 | 1.8331919 | 157.51703 | 573501.3 | -115975 |
| 0.3 | 2.64E-05 | 133.9177 | 6.415E-08 | 0.110353 | 1.7112065 | 162.20384 | 569743.1 | -119733 |
| 0.35 | 2.58E-05 | 137.078 | 6.297E-08 | 0.1003502 | 1.5912447 | 167.15842 | 565818.7 | -123657 |
| 0.4 | 2.52E-05 | 140.3764 | 6.174E-08 | 0.0908226 | 1.4734467 | 172.40017 | 561733.2 | -127743 |
| 0.45 | 2.46E-05 | 143.8228 | 6.047E-08 | 0.0817701 | 1.3579661 | 177.94785 | 557499.9 | -131976 |
| 0.5 | 2.4E-05 | 147.4281 | 5.913E-08 | 0.0731928 | 1.2449714 | 183.81799 | 553144.2 | -136332 |
| 0.55 | 2.34E-05 | 151.2039 | 5.77E-08 | 0.0650905 | 1.1346476 | 190.02212 | 548709.2 | -140767 |
| 0.6 | 2.29E-05 | 155.1635 | 5.618E-08 | 0.0574634 | 1.0271989 | 196.56217 | 544265.9 | -145210 |
| 0.65 | 2.23E-05 | 159.3213 | 5.451E-08 | 0.0503115 | 0.9228506 | 203.42289 | 539926.8 | -149549 |
| 0.7 | 2.17E-05 | 163.6932 | 5.268E-08 | 0.0436346 | 0.8218522 | 210.55959 | 535868.2 | -153608 |
| 0.75 | 2.11E-05 | 168.297 | 5.06E-08 | 0.0374329 | 0.7244808 | 217.87818 | 532365 | -157111 |
| 0.8 | 2.05E-05 | 173.1524 | 4.822E-08 | 0.0317063 | 0.6310448 | 225.20366 | 529845.2 | -159631 |
| 0.85 | 2E-05 | 178.2814 | 4.54E-08 | 0.0264549 | 0.5418888 | 232.23129 | 528977.5 | -160498 |
| 0.9 | 1.94E-05 | 183.7089 | 4.198E-08 | 0.0216786 | 0.4573987 | 238.45507 | 530812.7 | -158663 |
| 0.95 | 1.88E-05 | 189.4624 | 3.768E-08 | 0.0173774 | 0.3780081 | 243.07302 | 537013.7 | -152462 |
| 1 | 1.82E-05 | 195.5734 | 3.21E-08 | 0.0135513 | 0.3042062 | 244.88609 | 550236.3 | -139240 |

PR:

| $y_{H2}$ | b (PR) | V~_0 | rKI (PR) | a(T) | q | V~(1) | P(1) | ΔP |
|---|---|---|---|---|---|---|---|---|
| 0 | 2.68E-05 | 130.2714 | 5.744E-08 | 0.2043238 | 3.115496 | 152.71522 | 589505.6 | -99970.4 |
| 0.05 | 2.63E-05 | 133.0505 | 5.658E-08 | 0.1910539 | 2.9711047 | 156.74644 | 586511.5 | -102964 |
| 0.1 | 2.58E-05 | 135.9348 | 5.572E-08 | 0.1782295 | 2.8279213 | 160.98297 | 583371.5 | -106105 |
| 0.15 | 2.52E-05 | 138.931 | 5.484E-08 | 0.1658506 | 2.6860207 | 165.44121 | 580078.1 | -109398 |
| 0.2 | 2.47E-05 | 142.0461 | 5.395E-08 | 0.1539171 | 2.5454843 | 170.13893 | 576624.9 | -112851 |
| 0.25 | 2.42E-05 | 145.288 | 5.303E-08 | 0.1424292 | 2.4064007 | 175.09532 | 573006.4 | -116470 |
| 0.3 | 2.37E-05 | 148.6652 | 5.21E-08 | 0.1313868 | 2.2688661 | 180.33084 | 569219.4 | -120257 |
| 0.35 | 2.31E-05 | 152.1867 | 5.114E-08 | 0.1207899 | 2.1329854 | 185.86691 | 565264 | -124212 |
| 0.4 | 2.26E-05 | 155.8627 | 5.015E-08 | 0.1106384 | 1.9988733 | 191.72537 | 561145.4 | -128331 |
| 0.45 | 2.21E-05 | 159.7042 | 4.911E-08 | 0.1009325 | 1.8666554 | 197.92738 | 556876.8 | -132599 |
| 0.5 | 2.16E-05 | 163.7233 | 4.802E-08 | 0.0916721 | 1.7364692 | 204.49162 | 552483.4 | -136993 |
| 0.55 | 2.11E-05 | 167.9331 | 4.687E-08 | 0.0828571 | 1.6084662 | 211.43113 | 548008.9 | -141467 |
| 0.6 | 2.05E-05 | 172.3484 | 4.563E-08 | 0.0744877 | 1.4828131 | 218.74816 | 543524.3 | -145952 |
| 0.65 | 2E-05 | 176.9853 | 4.428E-08 | 0.0665638 | 1.3596941 | 226.42566 | 539142.9 | -150333 |
| 0.7 | 1.95E-05 | 181.8616 | 4.278E-08 | 0.0590853 | 1.2393131 | 234.41338 | 535042.2 | -154434 |
| 0.75 | 1.9E-05 | 186.9972 | 4.11E-08 | 0.0520524 | 1.1218967 | 242.60541 | 531499.1 | -157977 |
| 0.8 | 1.84E-05 | 192.4141 | 3.916E-08 | 0.045465 | 1.0076969 | 250.8046 | 528944.7 | -160531 |
| 0.85 | 1.79E-05 | 198.1371 | 3.687E-08 | 0.039323 | 0.8969953 | 258.66749 | 528053.1 | -161423 |
| 0.9 | 1.74E-05 | 204.1937 | 3.409E-08 | 0.0336266 | 0.7901074 | 265.62389 | 529883.3 | -159593 |
| 0.95 | 1.69E-05 | 210.6149 | 3.061E-08 | 0.0283757 | 0.6873875 | 270.77066 | 536110.7 | -153365 |
| 1 | 1.64E-05 | 217.4358 | 2.607E-08 | 0.0235702 | 0.5892356 | 272.75963 | 549410.2 | -140066 |

The above results demonstrate that the pipeline pressure drop exhibits a minimum at $y_{H_2} = 0.85$ and is a monotonically decreasing and increasing function of $y_{H_2}$ before and after $y_{H_2} = 0.85$. This is illustrated in Figure 7, which also demonstrates differences in pressure drop predictions of up to 3.045% between IGEOS and GCEOS. For $y_{H_2} = 0$, the IGEOS pressure drop prediction is identical to the baseline case in [BSL 15.4-2 example p.464-465], while the GCEOS pressure drop predictions for $y_{H_2} = 0$ are different at about 0.147% to 1.61%.

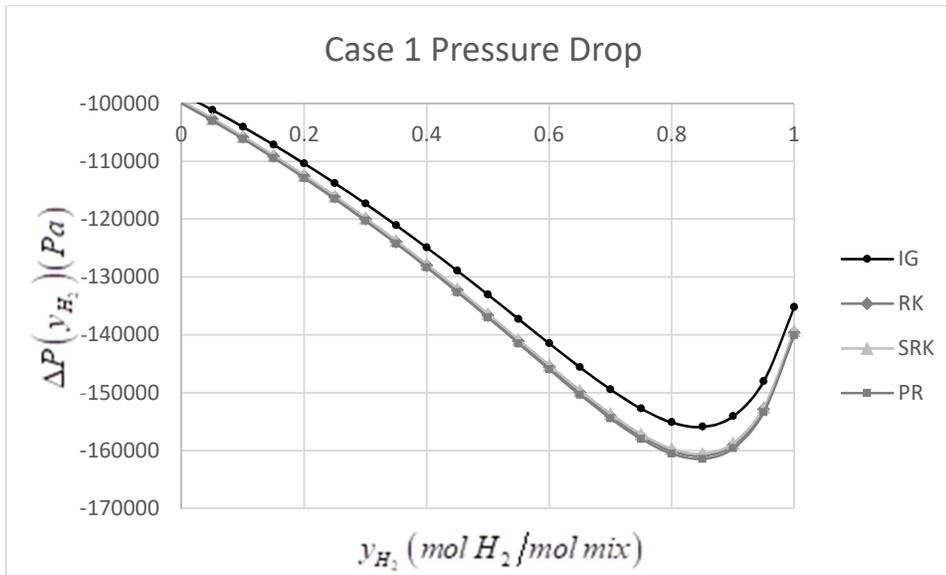

Figure 7: $\left(y_{H_2}, \Delta P(y_{H_2})\right)$ relations for Case 1 CH$_4$/H$_2$ mixture, IGEOS, GCEOS

**NG/H2 Case Study**

For this case study we employ information provided by a natural gas (NG) pipeline owning company regarding a realistic NG composition and NG pipeline and NG feed data, [Com24]. NG contains 8 species and has the following composition (in mole fraction $\{y_i\}_{i=1}^{N}$ terms): 95.124%

CH$_4$, 1.438% N$_2$, 0.530% CO$_2$, 2.721% C$_2$H$_6$, 0.161% C$_3$H$_8$, 0.011% i-C$_4$H$_{10}$, 0.012% n-C$_4$H$_{10}$, 0.003% n-C$_6$H$_{14}$.

The pipeline NG gas inlet pressure and velocity are:

$$P_0 = 734.56\, psia = 5064612.92\, Pa,\ v_0 = 258.11 \frac{m}{s}$$

The pipeline NG gas temperature is: $T = 83.35°F = 301.483K$

The pipeline dimensions are:

$$\begin{cases} D_p = 33.162\, in. = 0.8423148m \\ L_p = 24.31\, mi. = 39121.903m \end{cases} \Rightarrow \begin{cases} A_p = \pi \frac{D_p^2}{4} = \pi \frac{(0.8423148m)^2}{4} = 0.5572355 m^2 \\ r_{LD} = \frac{D_p}{L_p} = \frac{0.8423148m}{39121.903m} = 2.15305 \cdot 10^{-5} \end{cases}$$

The pipeline NG gas inlet mass flow rate is: $\dot{m} = 102.36 \frac{kg\, NG}{s}$

Then, the pipeline NG gas inlet mass density is:

$$\rho_{0,NG} = \frac{\dot{m}}{v_0 A_p} = \frac{102.36 \left(\frac{kg\, NG}{s}\right)}{258.11 \left(\frac{m}{s}\right) 0.5572355 (m^2)} = 0.7116831 \left(\frac{kg\, NG}{m^3\, NG}\right)$$

For the provided NG composition of 8 species, the NG molar mass is

$$M_{NG} \left(\frac{kg\, NG}{mol\, NG}\right) = 16.8019915 \cdot 10^{-3} \left(\frac{kg\, NG}{mol\, NG}\right)$$

Then, the pipeline NG gas inlet molar volume is

$$V_0 \left(\frac{m^3\, NG}{mol\, NG}\right) = \frac{M_{NG}\left(\frac{kg\, NG}{mol\, NG}\right)}{\rho_{0,NG}\left(\frac{kg\, NG}{m^3\, NG}\right)} = \frac{16.8019915 \cdot 10^{-3}}{0.7116831} \left(\frac{m^3\, NG}{mol\, NG}\right) = 0.023609 \left(\frac{m^3\, NG}{mol\, NG}\right).$$

The above information will help determine which of the IGEOS, GCEOS model molar volume predictions at the pipeline entrance are closest to the above molar volume calculated by the provided inlet data. In creating the NG/ H$_2$ blend composition ((in mole fraction $\{y_i\}_{i=1}^N$ terms), it is considered that the mole fraction ratios of the 8 species in NG are kept constant for any NG/ H$_2$ blend. The mass flowrates for blends of NG/H$_2$ are determined based on the Higher Heating Value (HHV) on a mass basis.

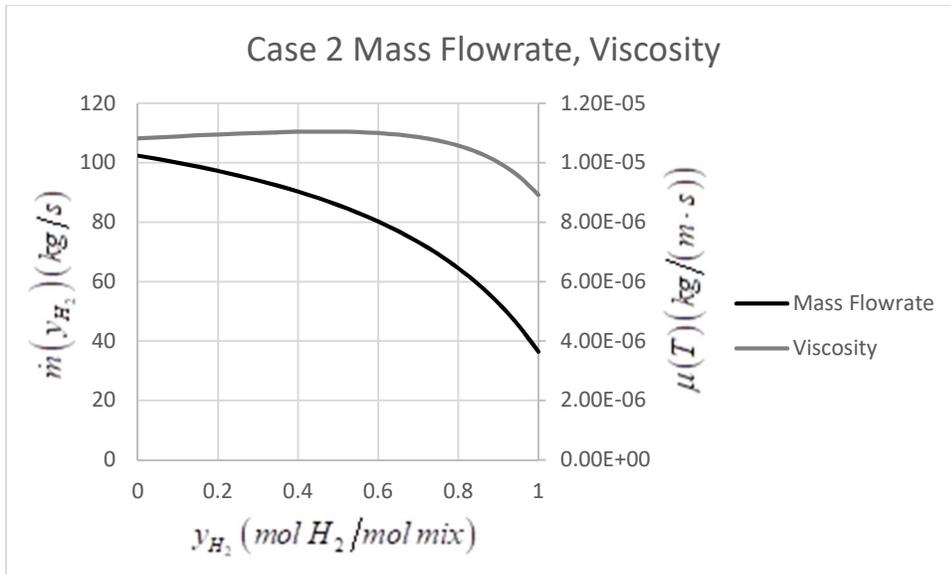

Figure 8: $\left(y_{H_2}, \dot{m}(y_{H_2})\right), \left(y_{H_2}, \mu(T)(y_{H_2})\right)$ relations for Case 2 NG/H$_2$ mixture

The resulting pressure drop model predictions for the considered EOS models and for various hydrogen mole fraction $y_{H_2}$ values are:

IG:

| y_H2 | b (IG) | V~_0 | rKI (IG) | V~(1) | P(1) | ΔP |
|---|---|---|---|---|---|---|
| 0 | 0.000349613 | 1.415594 | 9.79325E-05 | 1.476607 | 4855343.564 | -209269.356 |
| 0.05 | 0.000342641 | 1.444401 | 9.61947E-05 | 1.508216 | 4850320.683 | -214292.2373 |
| 0.1 | 0.000335668 | 1.474404 | 9.44312E-05 | 1.541201 | 4845110.383 | -219502.5367 |
| 0.15 | 0.000328695 | 1.505681 | 9.26376E-05 | 1.575650 | 4839712.426 | -224900.4943 |
| 0.2 | 0.000321723 | 1.538313 | 9.08086E-05 | 1.611657 | 4834129.92 | -230482.9996 |
| 0.25 | 0.00031475 | 1.572391 | 8.89377E-05 | 1.649325 | 4828370.715 | -236242.205 |
| 0.3 | 0.000307778 | 1.608013 | 8.70172E-05 | 1.688761 | 4822449.316 | -242163.6041 |
| 0.35 | 0.000300805 | 1.645286 | 8.50375E-05 | 1.730080 | 4816389.566 | -248223.3541 |
| 0.4 | 0.000293832 | 1.684329 | 8.29863E-05 | 1.773403 | 4810228.402 | -254384.5178 |
| 0.45 | 0.00028686 | 1.725269 | 8.08488E-05 | 1.818856 | 4804021.175 | -260591.7448 |
| 0.5 | 0.000279887 | 1.768249 | 7.86057E-05 | 1.866565 | 4797849.249 | -266763.6713 |
| 0.55 | 0.000272915 | 1.813426 | 7.62326E-05 | 1.916658 | 4791830.98 | -272781.9404 |
| 0.6 | 0.000265942 | 1.860971 | 7.36977E-05 | 1.969249 | 4786137.78 | -278475.1401 |
| 0.65 | 0.000258969 | 1.911077 | 7.09595E-05 | 2.024436 | 4781017.941 | -283594.9789 |
| 0.7 | 0.000251997 | 1.963955 | 6.79625E-05 | 2.082274 | 4776832.536 | -287780.3842 |
| 0.75 | 0.000245024 | 2.019843 | 6.46317E-05 | 2.142749 | 4774110.487 | -290502.4329 |
| 0.8 | 0.000238052 | 2.079005 | 6.08638E-05 | 2.205731 | 4773634.751 | -290978.1692 |
| 0.85 | 0.000231079 | 2.141737 | 5.65133E-05 | 2.270886 | 4776580.312 | -288032.6077 |
| 0.9 | 0.000224106 | 2.208372 | 5.13706E-05 | 2.337546 | 4784741.086 | -279871.8343 |
| 0.95 | 0.000217134 | 2.279288 | 4.51258E-05 | 2.404481 | 4800915.07 | -263697.8505 |
| 1 | 0.000210161 | 2.354909 | 3.73046E-05 | 2.469508 | 4829586.018 | -235026.9023 |

RK:

| y_H2 | b (RK) | V~_0 | rKI (RK) | a(T) | q | V~(1) | P(1) | ΔP |
|---|---|---|---|---|---|---|---|---|
| 0 | 3.029E-05 | 14.895372 | 7.351E-07 | 0.193469 | 2.548191 | 15.529376 | 4873876 | -190736.9 |
| 0.05 | 2.9686E-05 | 15.390995 | 7.221E-07 | 0.178416 | 2.397748 | 16.062629 | 4866825 | -197787.9 |
| 0.1 | 2.9082E-05 | 15.894320 | 7.088E-07 | 0.163973 | 2.249417 | 16.605836 | 4859648 | -204964.8 |
| 0.15 | 2.8478E-05 | 16.406639 | 6.954E-07 | 0.150139 | 2.103333 | 17.160430 | 4852341 | -212272.3 |
| 0.2 | 2.7874E-05 | 16.929239 | 6.817E-07 | 0.136915 | 1.95964 | 17.727840 | 4844902 | -219710.7 |
| 0.25 | 2.727E-05 | 17.463426 | 6.676E-07 | 0.124300 | 1.818499 | 18.309504 | 4837339 | -227274.3 |
| 0.3 | 2.6666E-05 | 18.010538 | 6.532E-07 | 0.112295 | 1.680083 | 18.906878 | 4829663 | -234950.1 |
| 0.35 | 2.6062E-05 | 18.571965 | 6.383E-07 | 0.100899 | 1.544581 | 19.521439 | 4821899 | -242714 |
| 0.4 | 2.5458E-05 | 19.149164 | 6.229E-07 | 0.090113 | 1.4122 | 20.154677 | 4814085 | -250528.1 |
| 0.45 | 2.4854E-05 | 19.743679 | 6.069E-07 | 0.079936 | 1.283169 | 20.808085 | 4806279 | -258334.4 |
| 0.5 | 2.4249E-05 | 20.357157 | 5.901E-07 | 0.070370 | 1.157738 | 21.483127 | 4798566 | -266047.1 |
| 0.55 | 2.3645E-05 | 20.991365 | 5.722E-07 | 0.061412 | 1.036182 | 22.181184 | 4791072 | -273541.3 |
| 0.6 | 2.3041E-05 | 21.648218 | 5.532E-07 | 0.053064 | 0.918806 | 22.903471 | 4783977 | -280636.2 |
| 0.65 | 2.2437E-05 | 22.329795 | 5.327E-07 | 0.045326 | 0.805949 | 23.650888 | 4777543 | -287070.3 |
| 0.7 | 2.1833E-05 | 23.038371 | 5.102E-07 | 0.038197 | 0.697984 | 24.423793 | 4772149 | -292464.1 |
| 0.75 | 2.1229E-05 | 23.776445 | 4.852E-07 | 0.031678 | 0.595331 | 25.221631 | 4768349 | -296263.4 |
| 0.8 | 2.0625E-05 | 24.546778 | 4.569E-07 | 0.025768 | 0.498454 | 26.042332 | 4766963 | -297649.8 |
| 0.85 | 2.0021E-05 | 25.352434 | 4.242E-07 | 0.020468 | 0.407879 | 26.881328 | 4769215 | -295398.1 |
| 0.9 | 1.9417E-05 | 26.196827 | 3.856E-07 | 0.015778 | 0.324191 | 27.729930 | 4776970 | -287642.6 |
| 0.95 | 1.8812E-05 | 27.083781 | 3.387E-07 | 0.011697 | 0.248056 | 28.572606 | 4793133 | -271480 |
| 1 | 1.8208E-05 | 28.017596 | 2.8E-07 | 0.008225 | 0.180225 | 29.382325 | 4822344 | -242268.6 |

SRK:

| y_H2 | b (SRK) | V~_0 | rKI (SRK) | a(T) | q | V~(1) | P(1) | ΔP |
|---|---|---|---|---|---|---|---|---|
| 0 | 3.029E-05 | 14.036102 | 7.351E-07 | 0.248463 | 3.272527 | 14.637047 | 4884854 | -179758.7 |
| 0.05 | 2.9686E-05 | 14.534375 | 7.221E-07 | 0.233569 | 3.138961 | 15.171251 | 4877814 | -186799.1 |
| 0.1 | 2.9082E-05 | 15.040334 | 7.088E-07 | 0.219136 | 3.00616 | 15.715412 | 4870646 | -193966.8 |
| 0.15 | 2.8478E-05 | 15.555469 | 6.954E-07 | 0.205163 | 2.874174 | 16.271159 | 4863343 | -201269.5 |
| 0.2 | 2.7874E-05 | 16.081245 | 6.817E-07 | 0.191650 | 2.743055 | 16.840101 | 4855902 | -208710.6 |
| 0.25 | 2.727E-05 | 16.619128 | 6.676E-07 | 0.178597 | 2.612861 | 17.423847 | 4848325 | -216287.5 |
| 0.3 | 2.6666E-05 | 17.170612 | 6.532E-07 | 0.166004 | 2.483655 | 18.024016 | 4840623 | -223990.1 |
| 0.35 | 2.6062E-05 | 17.737238 | 6.383E-07 | 0.153872 | 2.355505 | 18.642252 | 4832815 | -231798.3 |
| 0.4 | 2.5458E-05 | 18.320619 | 6.229E-07 | 0.142200 | 2.228487 | 19.280217 | 4824935 | -239677.8 |
| 0.45 | 2.4854E-05 | 18.922455 | 6.069E-07 | 0.130989 | 2.102683 | 19.939589 | 4817038 | -247575.2 |
| 0.5 | 2.4249E-05 | 19.544558 | 5.901E-07 | 0.120238 | 1.978184 | 20.622029 | 4809203 | -255410.2 |
| 0.55 | 2.3645E-05 | 20.188874 | 5.722E-07 | 0.109947 | 1.85509 | 21.329140 | 4801549 | -263063.9 |
| 0.6 | 2.3041E-05 | 20.857507 | 5.532E-07 | 0.100116 | 1.733512 | 22.062380 | 4794250 | -270363.3 |
| 0.65 | 2.2437E-05 | 21.552744 | 5.327E-07 | 0.090746 | 1.613571 | 22.822934 | 4787557 | -277055.9 |
| 0.7 | 2.1833E-05 | 22.277092 | 5.102E-07 | 0.081836 | 1.495405 | 23.611482 | 4781839 | -282773.8 |
| 0.75 | 2.1229E-05 | 23.033304 | 4.852E-07 | 0.073386 | 1.379163 | 24.427850 | 4777636 | -286976.7 |

| y_H2 | b (SRK) | V~_0 | rKI (SRK) | a(T) | q | V~(1) | P(1) | ΔP |
|---|---|---|---|---|---|---|---|---|
| 0.8 | 2.0625E-05 | 23.824429 | 4.569E-07 | 0.065397 | 1.265016 | 25.270415 | 4775749 | -288863.9 |
| 0.85 | 2.0021E-05 | 24.653850 | 4.242E-07 | 0.057868 | 1.153154 | 26.135140 | 4777380 | -287232.8 |
| 0.9 | 1.9417E-05 | 25.525348 | 3.856E-07 | 0.050799 | 1.043788 | 27.013973 | 4784366 | -280246.7 |
| 0.95 | 1.8812E-05 | 26.443161 | 3.387E-07 | 0.044191 | 0.93716 | 27.892147 | 4799573 | -265039.6 |
| 1 | 1.8208E-05 | 27.412066 | 2.8E-07 | 0.038043 | 0.833543 | 28.743544 | 4827594 | -237018.9 |

PR:

| y_H2 | b (SRK) | V~_0 | rKI (SRK) | a(T) | q | V~(1) | P(1) | ΔP |
|---|---|---|---|---|---|---|---|---|
| 0 | 2.72E-05 | 15.393815 | 5.928E-07 | 0.263982 | 3.871987 | 16.048250 | 4887607 | -177006.3 |
| 0.05 | 2.6657E-05 | 15.978765 | 5.823E-07 | 0.246719 | 3.692424 | 16.674327 | 4880229 | -184384.1 |
| 0.1 | 2.6115E-05 | 16.570250 | 5.716E-07 | 0.230040 | 3.514317 | 17.309481 | 4872744 | -191868.6 |
| 0.15 | 2.5573E-05 | 17.170171 | 5.607E-07 | 0.213945 | 3.337761 | 17.955770 | 4865143 | -199470.2 |
| 0.2 | 2.503E-05 | 17.780343 | 5.496E-07 | 0.198433 | 3.162855 | 18.615170 | 4857419 | -207194.1 |
| 0.25 | 2.4488E-05 | 18.402542 | 5.383E-07 | 0.183505 | 2.989709 | 19.289616 | 4849574 | -215039 |
| 0.3 | 2.3945E-05 | 19.038540 | 5.267E-07 | 0.169160 | 2.818443 | 19.981019 | 4841617 | -222996 |
| 0.35 | 2.3403E-05 | 19.690137 | 5.147E-07 | 0.155400 | 2.649188 | 20.691289 | 4833568 | -231045.4 |
| 0.4 | 2.286E-05 | 20.359184 | 5.023E-07 | 0.142223 | 2.482086 | 21.422336 | 4825460 | -239153.3 |
| 0.45 | 2.2318E-05 | 21.047613 | 4.894E-07 | 0.129629 | 2.317296 | 22.176059 | 4817347 | -247266.3 |
| 0.5 | 2.1775E-05 | 21.757463 | 4.758E-07 | 0.117620 | 2.154988 | 22.954326 | 4809309 | -255303.5 |
| 0.55 | 2.1233E-05 | 22.490905 | 4.614E-07 | 0.106194 | 1.995355 | 23.758917 | 4801467 | -263145.6 |
| 0.6 | 2.069E-05 | 23.250269 | 4.461E-07 | 0.095352 | 1.838606 | 24.591436 | 4793995 | -270618.1 |
| 0.65 | 2.0148E-05 | 24.038076 | 4.295E-07 | 0.085093 | 1.684974 | 25.453156 | 4787146 | -277467.2 |
| 0.7 | 1.9605E-05 | 24.857074 | 4.114E-07 | 0.075418 | 1.534718 | 26.344776 | 4781290 | -283322.5 |
| 0.75 | 1.9063E-05 | 25.710272 | 3.912E-07 | 0.066327 | 1.388126 | 27.266011 | 4776972 | -287640.8 |
| 0.8 | 1.852E-05 | 26.600986 | 3.684E-07 | 0.057820 | 1.24552 | 28.214940 | 4774996 | -289617.4 |
| 0.85 | 1.7978E-05 | 27.532892 | 3.421E-07 | 0.049896 | 1.107261 | 29.186920 | 4776568 | -288044.6 |
| 0.9 | 1.7435E-05 | 28.510081 | 3.109E-07 | 0.042556 | 0.973755 | 30.172795 | 4783534 | -281078.6 |
| 0.95 | 1.6893E-05 | 29.537134 | 2.731E-07 | 0.035799 | 0.845459 | 31.155882 | 4798768 | -265844.6 |
| 1 | 1.6351E-05 | 30.619202 | 2.258E-07 | 0.029626 | 0.722892 | 32.106798 | 4826875 | -237737.7 |

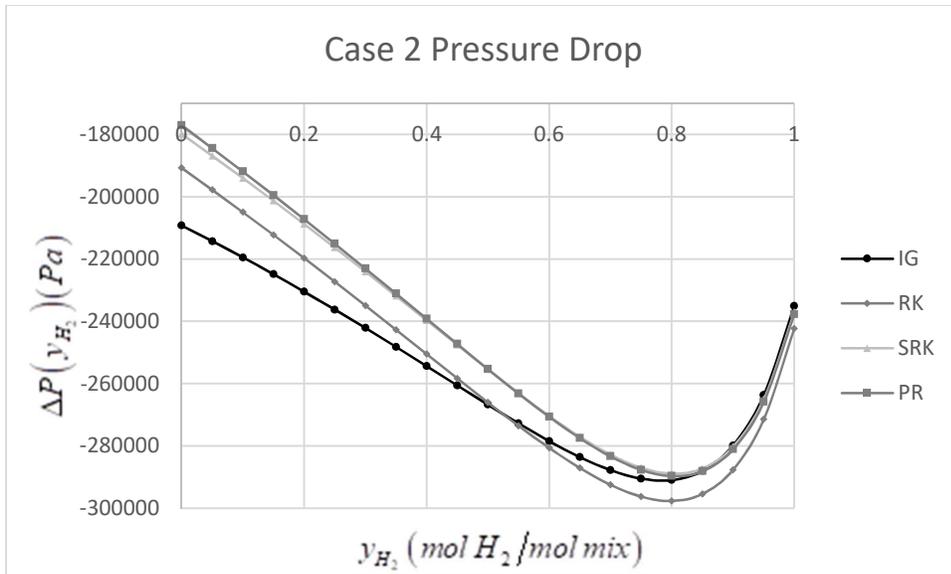

Figure 9: $\left(y_{H_2}, \Delta P(y_{H_2})\right)$ relations for Case 2 NG/H$_2$ mixture, IGEOS, GCEOS

**Conclusions:**

In this work, we presented a novel, dimensionless model that results in a dimensionless algebraic equation parametrized with only three dimensionless temperature, geometry, and flow related parameters $r_{KI}, q, r_{LD}$, thus providing the potential for parametric studies in a parameter space of reduced dimension compared to the parameter space of dimensional models. We quantified the pressure drop dependence on hydrogen mole fraction of a binary, methane hydrogen mixture, and a natural gas hydrogen mixture with a real life natural gas composition containing eight species using the IG and nonideal GCEOS, for steady state, isothermal, compressible flow through a straight, horizontal pipeline. Our calculations showed excellent agreement with simulations and calculations from prior works focusing on pure methane and methane-hydrogen blends.

%29%2B1%2F%282*a*x%5E2%29%22&assumption=%7B%22FP%22%2C+%22AsymptoticLimitCalculator%22%2C+%22limit%22%7D+-%3E+%22positive%22]

[Zha24] Zhang, B., Xu, N., Zhang, H., Qiu, R., Wei, X., Wang, Z., & Liang, Y. Influence of hydrogen blending on the operation of natural gas pipeline network considering the compressor power optimization. *Applied Energy*, *358*, 122594, (2024)

**Acknowledgement**

Financial support through the California Energy Commission Grant PIR-22-003. "Pilot Testing and Assessment of Safety and Integrity of Targeted Hydrogen Blending in Gas Infrastructure for Decarbonization" is gratefully acknowledged.

**Appendix**
A.1 Proof of Theorem
a. IGEOS:

$$M\left(\frac{4\dot{m}}{M\cdot\pi\cdot D_p^2}\right)^2 dV + M\left[\frac{2f}{D_p}\left(\frac{4\dot{m}}{M\cdot\pi\cdot D_p^2}\right)^2 V\right]dl + dP = 0 \Rightarrow$$

$$M\left(\frac{4\dot{m}}{M\cdot\pi\cdot D_p^2}\right)^2 \frac{b^{IG}}{RT}dV + M\left[\frac{2f}{D_p}\left(\frac{4\dot{m}}{M\cdot\pi\cdot D_p^2}\right)^2 V\right]\frac{b^{IG}}{RT}dl + \frac{b^{IG}}{RT}dP = 0 \underset{\{y_i\}_{i=1}^N=constant}{\overset{T=constant}{\Rightarrow}}$$

$$M\left(\frac{4\dot{m}}{M\cdot\pi\cdot D_p^2}\right)^2 \frac{b^{IG}}{RT}dV + M\left[\frac{2f}{D_p}\left(\frac{4\dot{m}}{M\cdot\pi\cdot D_p^2}\right)^2 V\right]\frac{b^{IG}}{RT}dl + d\left(\frac{Pb^{IG}}{RT}\right) = 0$$

Using the dimensionless variables $\beta^{IG} \triangleq \frac{Pb^{IG}}{RT}$, $\tilde{V}^{IG} \triangleq \frac{V}{b^{IG}}$ employed to create the DIGEOS $\beta^{IG} = \frac{1}{\tilde{V}^{IG}}$, then yields the mechanical energy balance dimensionless differential form

$$\frac{Mb^{IG}}{RT}\left(\frac{4\dot{m}}{M\cdot\pi\cdot D_p^2}\right)^2 d\left(b^{IG}\tilde{V}^{IG}\right) + \frac{Mb^{IG}}{RT}\left[\frac{2f}{D_p}\left(\frac{4\dot{m}}{M\cdot\pi\cdot D_p^2}\right)^2 b^{IG}\tilde{V}^{IG}\right]d\left(L_p\tilde{l}\right) + d\beta^{IG} = 0 \underset{\{y_i\}_{i=1}^N=constant}{\overset{L_p=constant}{\Rightarrow}}$$

$$\frac{M\left(b^{IG}\right)^2}{RT}\left(\frac{4\dot{m}}{M\cdot\pi\cdot D_p^2}\right)^2 d\tilde{V}^{IG} + \frac{M\left(b^{IG}\right)^2}{RT}\left[\frac{2fL_p}{D_p}\left(\frac{4\dot{m}}{M\cdot\pi\cdot D_p^2}\right)^2 \tilde{V}^{IG}\right]d\tilde{l} + d\left(\frac{1}{\tilde{V}^{IG}}\right) = 0 \Rightarrow$$

$$\left[\frac{M\left(b^{IG}\right)^2}{RT}\left(\frac{4\dot{m}}{M\cdot\pi\cdot D_p^2}\right)^2 - \frac{1}{\left(\tilde{V}^{IG}\right)^2}\right]d\tilde{V}^{IG} + \frac{M\left(b^{IG}\right)^2}{RT}\left[\frac{2fL_p}{D_p}\left(\frac{4\dot{m}}{M\cdot\pi\cdot D_p^2}\right)^2 \tilde{V}^{IG}\right]d\tilde{l} = 0 \Rightarrow$$

$$\dfrac{-\left[\dfrac{M\left(b^{IG}\right)^2}{RT}\left(\dfrac{4\dot{m}}{M\cdot\pi\cdot D_p^{\,2}}\right)^2-\dfrac{1}{\left(\tilde{V}^{IG}\right)^2}\right]}{\dfrac{M\left(b^{IG}\right)^2}{RT}\left(\dfrac{2fL_p}{D_p}\left(\dfrac{4\dot{m}}{M\cdot\pi\cdot D_p^{\,2}}\right)^2\tilde{V}^{IG}\right)}d\tilde{V}^{IG}=d\tilde{l}\Rightarrow\dfrac{-\left[r_{KI}^{IG}-\dfrac{1}{\left(\tilde{V}^{IG}\right)^2}\right]}{\dfrac{2f}{r_{LD}}r_{KI}^{IG}\tilde{V}^{IG}}d\tilde{V}^{IG}=d\tilde{l}\Rightarrow$$

$$\int_{\tilde{V}_0^{IG}}^{\tilde{V}^{IG}(1)}\left[\dfrac{-r_{LD}}{2f}\dfrac{1}{\tilde{V}^{IG}}+\dfrac{r_{LD}}{2fr_{KI}^{IG}}\left[\dfrac{1}{\left(\tilde{V}^{IG}\right)^3}\right]\right]d\tilde{V}^{IG}=\int_0^1 d\tilde{l}=1\Rightarrow$$

$$\left[-\left[\ln\left(\tilde{V}^{IG}(1)\right)-\ln\left(\tilde{V}_0^{IG}\right)\right]+\dfrac{1}{r_{KI}^{IG}}\left[-\dfrac{1}{2\left(\tilde{V}^{IG}(1)\right)^2}+\dfrac{1}{2\left(\tilde{V}_0^{IG}\right)^2}\right]\right]=\dfrac{2f}{r_{LD}}$$

Given the definitions

$$\gamma_1\left(\tilde{V}^{IG}\right)\triangleq -\ln\left(\tilde{V}^{IG}\right),\ \gamma_2^{IG}\left(\tilde{V}^{IG}\right)\triangleq\dfrac{-1}{2\left(\tilde{V}^{IG}\right)^2},\ \gamma^{IG}\left(\tilde{V}^{IG}\right)\triangleq \gamma_1\left(\tilde{V}^{IG}\right)+\dfrac{1}{r_{KI}^{IG}}\gamma_2^{IG}\left(\tilde{V}^{IG}\right)$$

the above equation becomes:

$$\boxed{\gamma^{IG}\left(\tilde{V}^{IG}(1)\right)-\gamma^{IG}\left(\tilde{V}_0^{IG}\right)\triangleq\left[\left[\gamma_1\left(\tilde{V}^{IG}(1)\right)+\dfrac{1}{r_{KI}^{IG}}\cdot\gamma_2^{IG}\left(\tilde{V}^{IG}(1)\right)\right]-\left[\gamma_1\left(\tilde{V}_0^{IG}\right)+\dfrac{1}{r_{KI}^{IG}}\cdot\gamma_2^{IG}\left(\tilde{V}_0^{IG}\right)\right]\right]=\dfrac{2f}{r_{LD}}}\ \text{ΟΕΔ}$$

b. GCEOS:

$$M\left(\dfrac{4\dot{m}}{M\cdot\pi\cdot D_p^{\,2}}\right)^2 dV+M\left(\dfrac{2f}{D_p}\left(\dfrac{4\dot{m}}{M\cdot\pi\cdot D_p^{\,2}}\right)^2 V\right)dl+dP=0\Rightarrow$$

$$M\left(\dfrac{4\dot{m}}{M\cdot\pi\cdot D_p^{\,2}}\right)^2\dfrac{b}{RT}dV+M\left(\dfrac{2f}{D_p}\left(\dfrac{4\dot{m}}{M\cdot\pi\cdot D_p^{\,2}}\right)^2 V\right)\dfrac{b}{RT}dl+\dfrac{b}{RT}dP=0\ \underset{\{y_i\}_{i=1}^N=constant}{\overset{T=constant}{\Rightarrow}}$$

$$M\left(\dfrac{4\dot{m}}{M\cdot\pi\cdot D_p^{\,2}}\right)^2\dfrac{b}{RT}dV+M\left(\dfrac{2f}{D_p}\left(\dfrac{4\dot{m}}{M\cdot\pi\cdot D_p^{\,2}}\right)^2 V\right)\dfrac{b}{RT}dl+d\left(\dfrac{Pb}{RT}\right)=0$$

and also using the dimensionless variables $\beta\triangleq\dfrac{Pb}{RT}>0,\ q\triangleq\dfrac{a(T)}{bRT}>0,\ \tilde{V}\triangleq\dfrac{V}{b}$ employed to create DGCEOS, then yields the following dimensionless differential form of the mechanical energy balance:

$$\dfrac{Mb}{RT}\left(\dfrac{4\dot{m}}{M\cdot\pi\cdot D_p^{\,2}}\right)^2 d\left(b\tilde{V}\right)+\dfrac{Mb}{RT}\left(\dfrac{2f}{D_p}\left(\dfrac{4\dot{m}}{M\cdot\pi\cdot D_p^{\,2}}\right)^2 b\tilde{V}\right)d\left(L_p\tilde{l}\right)+d\beta=0\ \underset{\{y_i\}_{i=1}^N=constant}{\overset{L_p=constant}{\Rightarrow}}$$

$$\dfrac{Mb^2}{RT}\left(\dfrac{4\dot{m}}{M\cdot\pi\cdot D_p^{\,2}}\right)^2 d\tilde{V}+\dfrac{Mb^2}{RT}\left(\dfrac{2fL_p}{D_p}\left(\dfrac{4\dot{m}}{M\cdot\pi\cdot D_p^{\,2}}\right)^2\tilde{V}\right)d\tilde{l}+\left[\dfrac{\partial\beta(q,\tilde{V})}{\partial\tilde{V}}d\tilde{V}+\dfrac{\partial\beta(q,\tilde{V})}{\partial q}dq\right]=0\ \overset{q=constant}{\Rightarrow}$$

$$\left[\frac{Mb^2}{RT}\left(\frac{4\dot{m}}{M\cdot\pi\cdot D_p^2}\right)^2 + \frac{\partial\beta(q,\tilde{V})}{\partial\tilde{V}}\right]d\tilde{V} + \frac{Mb^2}{RT}\left(\frac{2fL_p}{D_p}\left(\frac{4\dot{m}}{M\cdot\pi\cdot D_p^2}\right)^2\tilde{V}\right)d\tilde{l} = 0 \Rightarrow$$

$$\frac{-\left[\frac{Mb^2}{RT}\left(\frac{4\dot{m}}{M\cdot\pi\cdot D_p^2}\right)^2 + \frac{\partial\beta(q,\tilde{V})}{\partial\tilde{V}}\right]}{\frac{Mb^2}{RT}\left(\frac{2fL_p}{D_p}\left(\frac{4\dot{m}}{M\cdot\pi\cdot D_p^2}\right)^2\tilde{V}\right)}d\tilde{V} = d\tilde{l} \Rightarrow$$

$$\frac{-\left[\frac{Mb^2}{RT}\left(\frac{4\dot{m}}{M\cdot\pi\cdot D_p^2}\right)^2 - \frac{1}{(\tilde{V}-1)^2} + \frac{q}{\sigma-\varepsilon}\left[\frac{1}{(\tilde{V}+\varepsilon)^2} - \frac{1}{(\tilde{V}+\sigma)^2}\right]\right]}{\frac{Mb^2}{RT}\left(\frac{2fL_p}{D_p}\left(\frac{4\dot{m}}{M\cdot\pi\cdot D_p^2}\right)^2\tilde{V}\right)}d\tilde{V} = d\tilde{l} \Rightarrow$$

$$\begin{cases}\dfrac{-\left[r_{KI} + \dfrac{-1}{(\tilde{V}-1)^2} + \dfrac{q}{\sigma-\varepsilon}\left[\dfrac{1}{(\tilde{V}+\varepsilon)^2} - \dfrac{1}{(\tilde{V}+\sigma)^2}\right]\right]}{\left(\dfrac{2fr_{KI}}{r_{LD}}\tilde{V}\right)}d\tilde{V} = d\tilde{l} \quad \text{if } \varepsilon\neq 0,\ \sigma\neq 0 \\[2em] \dfrac{-\left[r_{KI} + \dfrac{-1}{(\tilde{V}-1)^2} + \dfrac{q}{\sigma}\left[\dfrac{1}{(\tilde{V})^2} - \dfrac{1}{(\tilde{V}+\sigma)^2}\right]\right]}{\left(\dfrac{2fr_{KI}}{r_{LD}}\tilde{V}\right)}d\tilde{V} = d\tilde{l} \quad \text{if } \varepsilon = 0,\ \sigma\neq 0\end{cases} \Rightarrow$$

$$\begin{cases}\left[\dfrac{-r_{KI}}{\left(\dfrac{2fr_{KI}}{r_{LD}}\tilde{V}\right)} + \dfrac{1}{(\tilde{V}-1)^2\left(\dfrac{2fr_{KI}}{r_{LD}}\tilde{V}\right)} - \dfrac{\dfrac{q}{\sigma-\varepsilon}}{(\tilde{V}+\varepsilon)^2\left(\dfrac{2fr_{KI}}{r_{LD}}\tilde{V}\right)} + \dfrac{\dfrac{q}{\sigma-\varepsilon}}{(\tilde{V}+\sigma)^2\left(\dfrac{2fr_{KI}}{r_{LD}}\tilde{V}\right)}\right]d\tilde{V} = d\tilde{l} \quad \text{if } \varepsilon\neq 0,\ \sigma\neq 0 \\[2em] \left[\dfrac{-r_{KI}}{\left(\dfrac{2fr_{KI}}{r_{LD}}\tilde{V}\right)} + \dfrac{1}{(\tilde{V}-1)^2\left(\dfrac{2fr_{KI}}{r_{LD}}\tilde{V}\right)} - \dfrac{\dfrac{q}{\sigma}}{(\tilde{V})^2\left(\dfrac{2fr_{KI}}{r_{LD}}\tilde{V}\right)} + \dfrac{\dfrac{q}{\sigma}}{(\tilde{V}+\sigma)^2\left(\dfrac{2fr_{KI}}{r_{LD}}\tilde{V}\right)}\right]d\tilde{V} = d\tilde{l} \quad \text{if } \varepsilon = 0,\ \sigma\neq 0\end{cases} \Rightarrow$$

Partial fraction expansion of the ratios that appear in the above differential form of the dimensionless mechanical energy balance yields:

$$\left\{\begin{array}{l}\left[\begin{array}{l}\dfrac{r_{LD}}{2f}\dfrac{-1}{\tilde{V}}+\dfrac{r_{LD}}{2fr_{KI}}\cdot\left[\dfrac{-\tilde{V}+2}{\left(\tilde{V}-1\right)^2}+\dfrac{1}{\tilde{V}}\right]-\dfrac{r_{LD}}{2fr_{KI}}\dfrac{q}{\sigma-\varepsilon}\left[\dfrac{-\tilde{V}-2\varepsilon}{\varepsilon^2\left(\tilde{V}+\varepsilon\right)^2}+\dfrac{1}{\varepsilon^2\tilde{V}}\right]+\\ +\dfrac{r_{LD}}{2fr_{KI}}\dfrac{q}{\sigma-\varepsilon}\left[\dfrac{-\tilde{V}-2\sigma}{\sigma^2\left(\tilde{V}+\sigma\right)^2}+\dfrac{1}{\sigma^2\tilde{V}}\right]\end{array}\right]d\tilde{V}=d\tilde{l}\ \ if\ \varepsilon\neq 0,\ \sigma\neq 0\\[2em] \left[\begin{array}{l}\dfrac{r_{LD}}{2f}\dfrac{-1}{\tilde{V}}+\dfrac{r_{LD}}{2fr_{KI}}\cdot\left[\dfrac{-\tilde{V}+2}{\left(\tilde{V}-1\right)^2}+\dfrac{1}{\tilde{V}}\right]-\dfrac{r_{LD}}{2fr_{KI}}\dfrac{q}{\sigma}\left[\dfrac{1}{\tilde{V}^3}\right]+\\ +\dfrac{r_{LD}}{2fr_{KI}}\dfrac{q}{\sigma}\left[\dfrac{-\tilde{V}-2\sigma}{\sigma^2\left(\tilde{V}+\sigma\right)^2}+\dfrac{1}{\sigma^2\tilde{V}}\right]\end{array}\right]d\tilde{V}=d\tilde{l}\ \ if\ \varepsilon=0,\ \sigma\neq 0\end{array}\right.$$

Given the definitions

$$\left\{\begin{array}{l}\gamma_1\left(\tilde{V}\right)\triangleq -ln\left(\tilde{V}\right),\ \gamma_2^{GC}\left(\tilde{V}\right)\triangleq ln\left(\dfrac{\tilde{V}}{\tilde{V}-1}\right)-\dfrac{1}{\left(\tilde{V}-1\right)},\\[1em] \gamma_3^{\varepsilon\sigma}\left(\tilde{V}\right)\triangleq \dfrac{-1}{\varepsilon^2\left(\sigma-\varepsilon\right)}\left[ln\left(\dfrac{\tilde{V}}{\varepsilon+\tilde{V}}\right)+\dfrac{\varepsilon}{\varepsilon+\tilde{V}}\right]+\dfrac{1}{\sigma^2\left(\sigma-\varepsilon\right)}\left[ln\left(\dfrac{\tilde{V}}{\sigma+\tilde{V}}\right)+\dfrac{\sigma}{\sigma+\tilde{V}}\right]\ if\ \varepsilon\neq 0,\ \sigma\neq 0\\[1em] \gamma_3^{0\sigma}\left(\tilde{V}\right)\triangleq \dfrac{1}{\sigma^2}\left[\dfrac{1}{\sigma}ln\left(\dfrac{\tilde{V}}{\sigma+\tilde{V}}\right)+\dfrac{1}{\sigma+\tilde{V}}\right]+\dfrac{1}{2\sigma\left(\tilde{V}\right)^2}\ if\ \varepsilon=0,\ \sigma\neq 0\\[1em] \gamma^{\varepsilon\sigma}\left(\tilde{V}\right)\triangleq \left[\gamma_1\left(\tilde{V}\right)+\dfrac{1}{r_{KI}}\cdot\gamma_2^{GC}\left(\tilde{V}\right)+\dfrac{q}{r_{KI}}\cdot\gamma_3^{\varepsilon\sigma}\left(\tilde{V}\right)\right]\ if\ \varepsilon\neq 0,\ \sigma\neq 0\\[1em] \gamma^{0\sigma}\left(\tilde{V}\right)\triangleq \left[\gamma_1\left(\tilde{V}\right)+\dfrac{1}{r_{KI}}\cdot\gamma_2^{GC}\left(\tilde{V}\right)+\dfrac{q}{r_{KI}}\cdot\gamma_3^{0\sigma}\left(\tilde{V}\right)\right]\ if\ \varepsilon=0,\ \sigma\neq 0\end{array}\right.,$$

the integration of the above differential form of the dimensionless mechanical energy balance along the dimensionless pipeline length then yields:

For $\varepsilon\neq 0,\ \sigma\neq 0$:

$$\int_{\tilde{V}_0}^{\tilde{V}(1)}\left[\begin{array}{l}\dfrac{-r_{LD}}{2f}\dfrac{1}{\tilde{V}}+\dfrac{r_{LD}}{2fr_{KI}}\cdot\left[\dfrac{-\tilde{V}+2}{\left(\tilde{V}-1\right)^2}+\dfrac{1}{\tilde{V}}\right]-\dfrac{r_{LD}}{2fr_{KI}}\dfrac{q}{\varepsilon^2\left(\sigma-\varepsilon\right)}\left[\dfrac{-\tilde{V}-2\varepsilon}{\left(\tilde{V}+\varepsilon\right)^2}+\dfrac{1}{\tilde{V}}\right]+\\ +\dfrac{r_{LD}}{2fr_{KI}}\dfrac{q}{\sigma^2\left(\sigma-\varepsilon\right)}\left[\dfrac{-\tilde{V}-2\sigma}{\left(\tilde{V}+\sigma\right)^2}+\dfrac{1}{\tilde{V}}\right]\end{array}\right]d\tilde{V}=\int_0^1 d\tilde{l}=1\underset{[Wol24b]}{\overset{[Wol24a]}{\Rightarrow}}$$

$$\left[\begin{array}{l} \dfrac{-r_{LD}}{2f}\left[\ln(\tilde{V}(1))-\ln(\tilde{V}_0)\right]+\dfrac{r_{LD}}{2fr_{KI}}\cdot\left[\left[-\ln(\tilde{V}(1)-1)-\dfrac{1}{(\tilde{V}(1)-1)}\right]-\left[-\ln(\tilde{V}_0-1)-\dfrac{1}{(\tilde{V}_0-1)}\right]+\ln(\tilde{V}(1))-\ln(\tilde{V}_0)\right]- \\ -\dfrac{r_{LD}}{2fr_{KI}}\dfrac{q}{\varepsilon^2(\sigma-\varepsilon)}\left[\left[\left[-\ln(\varepsilon+\tilde{V}(1))+\dfrac{\varepsilon}{\varepsilon+\tilde{V}(1)}\right]-\left[-\ln(\varepsilon+\tilde{V}_0)+\dfrac{\varepsilon}{\varepsilon+\tilde{V}_0}\right]\right]+\left[\ln(\tilde{V}(1))-\ln(\tilde{V}_0)\right]\right]+ \\ +\dfrac{r_{LD}}{2fr_{KI}}\dfrac{q}{\sigma^2(\sigma-\varepsilon)}\left[\left[\left[-\ln(\sigma+\tilde{V}(1))+\dfrac{\sigma}{\sigma+\tilde{V}(1)}\right]-\left[-\ln(\sigma+\tilde{V}_0)+\dfrac{\sigma}{\sigma+\tilde{V}_0}\right]\right]+\left[\ln(\tilde{V}(1))-\ln(\tilde{V}_0)\right]\right] \end{array}\right]=1\Rightarrow$$

$$\left[\begin{array}{l} -\left[\ln(\tilde{V}(1))-\ln(\tilde{V}_0)\right]+\dfrac{1}{r_{KI}}\cdot\left[\left[-\ln(\tilde{V}(1)-1)-\dfrac{1}{(\tilde{V}(1)-1)}\right]-\left[-\ln(\tilde{V}_0-1)-\dfrac{1}{(\tilde{V}_0-1)}\right]+\ln(\tilde{V}(1))-\ln(\tilde{V}_0)\right]- \\ -\dfrac{1}{r_{KI}}\dfrac{q}{\varepsilon^2(\sigma-\varepsilon)}\left[\left[\left[-\ln(\varepsilon+\tilde{V}(1))+\dfrac{\varepsilon}{\varepsilon+\tilde{V}(1)}\right]-\left[-\ln(\varepsilon+\tilde{V}_0)+\dfrac{\varepsilon}{\varepsilon+\tilde{V}_0}\right]\right]+\left[\ln(\tilde{V}(1))-\ln(\tilde{V}_0)\right]\right]+ \\ +\dfrac{1}{r_{KI}}\dfrac{q}{\sigma^2(\sigma-\varepsilon)}\left[\left[\left[-\ln(\sigma+\tilde{V}(1))+\dfrac{\sigma}{\sigma+\tilde{V}(1)}\right]-\left[-\ln(\sigma+\tilde{V}_0)+\dfrac{\sigma}{\sigma+\tilde{V}_0}\right]\right]+\left[\ln(\tilde{V}(1))-\ln(\tilde{V}_0)\right]\right] \end{array}\right]=\dfrac{2f}{r_{LD}}\Rightarrow$$

$$\boxed{\gamma^{\varepsilon\sigma}(\tilde{V}(1))-\gamma^{\varepsilon\sigma}(\tilde{V}_0)\triangleq\left[\begin{array}{l}\left[\gamma_1(\tilde{V}(1))+\dfrac{1}{r_{KI}}\cdot\gamma_2^{GC}(\tilde{V}(1))+\dfrac{q}{r_{KI}}\cdot\gamma_3^{\varepsilon\sigma}(\tilde{V}(1))\right]- \\ -\left[\gamma_1(\tilde{V}_0)+\dfrac{1}{r_{KI}}\cdot\gamma_2^{GC}(\tilde{V}_0)+\dfrac{q}{r_{KI}}\cdot\gamma_3^{\varepsilon\sigma}(\tilde{V}_0)\right]\end{array}\right]=\dfrac{2f}{r_{LD}}\text{ if }\varepsilon\ne 0,\sigma\ne 0}$$

For $\varepsilon=0,\sigma\ne 0$:

$$\int_{\tilde{V}_0}^{\tilde{V}(1)}\left[\dfrac{-r_{LD}}{2f}\dfrac{1}{\tilde{V}}+\dfrac{r_{LD}}{2fr_{KI}}\cdot\left[\dfrac{-\tilde{V}+2}{(\tilde{V}-1)^2}+\dfrac{1}{\tilde{V}}\right]-\dfrac{r_{LD}}{2fr_{KI}}\dfrac{q}{\sigma}\left[\dfrac{1}{\tilde{V}^3}\right]+\dfrac{r_{LD}}{2fr_{KI}}\dfrac{q}{\sigma^3}\left[\dfrac{-\tilde{V}-2\sigma}{(\tilde{V}+\sigma)^2}+\dfrac{1}{\tilde{V}}\right]\right]d\tilde{V}=\int_0^1 d\tilde{l}=1\Rightarrow$$

$$\left[\begin{array}{l} -\left[\ln(\tilde{V}(1))-\ln(\tilde{V}_0)\right]+\dfrac{1}{r_{KI}}\cdot\left[\left[-\ln(\tilde{V}(1)-1)-\dfrac{1}{(\tilde{V}(1)-1)}\right]-\left[-\ln(\tilde{V}_0-1)-\dfrac{1}{(\tilde{V}_0-1)}\right]+\ln(\tilde{V}(1))-\ln(\tilde{V}_0)\right]- \\ -\dfrac{1}{r_{KI}}\dfrac{q}{\sigma}\left[-\dfrac{1}{2(\tilde{V}(1))^2}+\dfrac{1}{2(\tilde{V}_0)^2}\right]+ \\ +\dfrac{1}{r_{KI}}\dfrac{q}{\sigma^3}\left[\left[\left[-\ln(\sigma+\tilde{V}(1))+\dfrac{\sigma}{\sigma+\tilde{V}(1)}\right]-\left[-\ln(\sigma+\tilde{V}_0)+\dfrac{\sigma}{\sigma+\tilde{V}_0}\right]\right]+\left[\ln(\tilde{V}(1))-\ln(\tilde{V}_0)\right]\right] \end{array}\right]=\dfrac{2f}{r_{LD}}\Rightarrow$$

$$\boxed{\gamma^{0\sigma}(\tilde{V}(1))-\gamma^{0\sigma}(\tilde{V}_0)\triangleq\left[\begin{array}{l}\left[\gamma_1(\tilde{V}(1))+\dfrac{1}{r_{KI}}\cdot\gamma_2^{GC}(\tilde{V}(1))+\dfrac{q}{r_{KI}}\cdot\gamma_3^{0\sigma}(\tilde{V}(1))\right]- \\ -\left[\gamma_1(\tilde{V}_0)+\dfrac{1}{r_{KI}}\cdot\gamma_2^{GC}(\tilde{V}_0)+\dfrac{q}{r_{KI}}\cdot\gamma_3^{0\sigma}(\tilde{V}_0)\right]\end{array}\right]=\dfrac{2f}{r_{LD}}\text{ if }\varepsilon=0,\sigma\ne 0}\text{ O.E.Δ.}$$

A.2. Case Study 1 CH$_4$-H$_2$ mix viscosity equation analysis:

$$\mu(T) = \frac{y_{H_2}\mu_{H_2}(T)}{y_{H_2}\Phi_{H_2,H_2}(T) + y_{CH_4}\Phi_{H_2,CH_4}(T)} + \frac{y_{CH_4}\mu_{CH_4}(T)}{y_{H_2}\Phi_{CH_4,H_2}(T) + y_{CH_4}\Phi_{CH_4,CH_4}(T)} \overset{\Phi_{H_2,H_2}(T)=1}{\underset{\Phi_{CH_4,CH_4}(T)=1}{\Rightarrow}}$$

$$\mu(T) = \frac{y_{H_2}\mu_{H_2}(T)}{y_{H_2} + (1-y_{H_2})\Phi_{H_2,CH_4}(T)} + \frac{(1-y_{H_2})\mu_{CH_4}(T)}{y_{H_2}\Phi_{CH_4,H_2}(T) + (1-y_{H_2})} \Rightarrow$$

$$\mu(T) = \frac{y_{H_2}\mu_{H_2}(T)}{\Phi_{H_2,CH_4}(T) + y_{H_2}(1-\Phi_{H_2,CH_4}(T))} + \frac{(1-y_{H_2})\mu_{CH_4}(T)}{1 + y_{H_2}(\Phi_{CH_4,H_2}(T) - 1)} \Rightarrow$$

$$\mu(T) = \frac{\begin{bmatrix} y_{H_2}\mu_{H_2}(T)\left[1 + y_{H_2}(\Phi_{CH_4,H_2}(T) - 1)\right] + \\ + (1-y_{H_2})\mu_{CH_4}(T)\left[\Phi_{H_2,CH_4}(T) + y_{H_2}(1-\Phi_{H_2,CH_4}(T))\right] \end{bmatrix}}{\left[\Phi_{H_2,CH_4}(T) + y_{H_2}(1-\Phi_{H_2,CH_4}(T))\right]\left[1 + y_{H_2}(\Phi_{CH_4,H_2}(T) - 1)\right]} \Rightarrow$$

$$\mu(T) = \frac{\left[\mu_{CH_4}(T)\Phi_{H_2,CH_4}(T)\right] + \begin{bmatrix} \mu_{H_2}(T) - \mu_{CH_4}(T)\Phi_{H_2,CH_4}(T) + \\ +\mu_{CH_4}(T)(1-\Phi_{H_2,CH_4}(T)) \end{bmatrix} y_{H_2} + \begin{bmatrix} \mu_{H_2}(T)(\Phi_{CH_4,H_2}(T) - 1) - \\ -\mu_{CH_4}(T)(1-\Phi_{H_2,CH_4}(T)) \end{bmatrix} y_{H_2}^2}{\left[\Phi_{H_2,CH_4}(T) + y_{H_2}(1-\Phi_{H_2,CH_4}(T))\right]\left[1 + y_{H_2}(\Phi_{CH_4,H_2}(T) - 1)\right]} \Rightarrow$$

$$\mu(T) = \frac{\left[\mu_{CH_4}(T)\Phi_{H_2,CH_4}(T)\right] + \begin{bmatrix} \mu_{H_2}(T) + \mu_{CH_4}(T) - \\ -2\mu_{CH_4}(T)\Phi_{H_2,CH_4}(T) \end{bmatrix} y_{H_2} + \begin{bmatrix} \mu_{H_2}(T)(\Phi_{CH_4,H_2}(T) - 1) + \\ +\mu_{CH_4}(T)(\Phi_{H_2,CH_4}(T) - 1) \end{bmatrix} y_{H_2}^2}{\left[\Phi_{H_2,CH_4}(T) + y_{H_2}(1-\Phi_{H_2,CH_4}(T))\right]\left[1 + y_{H_2}(\Phi_{CH_4,H_2}(T) - 1)\right]}$$

Considering $\mu(T): y_{H_2} \to \mu(T)(y_{H_2})$ as a function of $y_{H_2}$, then yields that the derivative

$\dfrac{d\mu(T)}{dy_{H_2}}: y_{H_2} \to \dfrac{d\mu(T)}{dy_{H_2}}(y_{H_2})$ is:

$$\frac{d\mu(T)}{dy_{H_2}}(y_{H_2}) = \frac{\begin{bmatrix} \begin{bmatrix} \mu_{H_2}(T) + \mu_{CH_4}(T) - \\ -2\mu_{CH_4}(T)\Phi_{H_2,CH_4}(T) \end{bmatrix}\Phi_{H_2,CH_4}(T) - \\ -[\mu_{CH_4}(T)\Phi_{H_2,CH_4}(T)](1-\Phi_{H_2,CH_4}(T)) - \\ -[\mu_{CH_4}(T)\Phi_{H_2,CH_4}(T)]\Phi_{H_2,CH_4}(T)(\Phi_{CH_4,H_2}(T)-1) \end{bmatrix} + \\ +2\begin{bmatrix} \begin{bmatrix} \mu_{H_2}(T)(\Phi_{CH_4,H_2}(T)-1) + \\ +\mu_{CH_4}(T)(\Phi_{H_2,CH_4}(T)-1) \end{bmatrix}\Phi_{H_2,CH_4}(T) - \\ -[\mu_{CH_4}(T)\Phi_{H_2,CH_4}(T)](1-\Phi_{H_2,CH_4}(T))(\Phi_{CH_4,H_2}(T)-1) \end{bmatrix} y_{H_2} + \\ +\begin{bmatrix} \begin{bmatrix} \mu_{H_2}(T)(\Phi_{CH_4,H_2}(T)-1) + \\ +\mu_{CH_4}(T)(\Phi_{H_2,CH_4}(T)-1) \end{bmatrix}(1-\Phi_{H_2,CH_4}(T)) + \\ +\begin{bmatrix} \mu_{H_2}(T)(\Phi_{CH_4,H_2}(T)-1) + \\ +\mu_{CH_4}(T)(\Phi_{H_2,CH_4}(T)-1) \end{bmatrix}\Phi_{H_2,CH_4}(T)(\Phi_{CH_4,H_2}(T)-1) - \\ -\begin{bmatrix} \mu_{H_2}(T) + \mu_{CH_4}(T) - \\ -2\mu_{CH_4}(T)\Phi_{H_2,CH_4}(T) \end{bmatrix}(1-\Phi_{H_2,CH_4}(T))(\Phi_{CH_4,H_2}(T)-1) \end{bmatrix} y_{H_2}^2}{\left[\Phi_{H_2,CH_4}(T) + y_{H_2}(1-\Phi_{H_2,CH_4}(T))\right]^2 \left[1 + y_{H_2}(\Phi_{CH_4,H_2}(T)-1)\right]^2}$$

For case study 1, substituting:

$$\begin{cases} \mu_{H_2}(T) = 8.9190 \cdot 10^{-6} \left(\frac{kg\,H_2}{m \cdot s}\right) \\ \mu_{CH_4}(T) = 1.0741 \cdot 10^{-5} \left(\frac{kg\,H_2}{m \cdot s}\right) \\ \Phi_{H_2,CH_4}(T) = \frac{\left[1 + (8.9190 \cdot 10^{-6}/1.0741 \cdot 10^{-5})^{1/2} \cdot (16.043 \cdot 10^{-3}/2.016 \cdot 10^{-3})^{1/4}\right]^2}{\left[8(1+(2.016 \cdot 10^{-3}/16.043 \cdot 10^{-3}))\right]^{1/2}} = 2.1338 \\ \Phi_{CH_4,H_2}(T) = \frac{\left[1 + (1.0741 \cdot 10^{-5}/8.9190 \cdot 10^{-6})^{1/2} \cdot (2.016 \cdot 10^{-3}/16.043 \cdot 10^{-3})^{1/4}\right]^2}{\left[8(1+(16.043 \cdot 10^{-3}/2.016 \cdot 10^{-3}))\right]^{1/2}} = 0.32292 \end{cases}$$

$$\frac{d\mu(T)}{dy_{H_2}}(y_{H_2}) = \frac{\begin{bmatrix} \begin{bmatrix} \begin{bmatrix} 8.9190 \cdot 10^{-6} + 1.0741 \cdot 10^{-5} - \\ -2 \cdot 1.0741 \cdot 10^{-5} \cdot 2.1338 \end{bmatrix} \cdot 2.1338 - \\ -\begin{bmatrix} 1.0741 \cdot 10^{-5} \cdot 2.1338 \end{bmatrix}(1-2.1338) - \\ -\begin{bmatrix} 1.0741 \cdot 10^{-5} \cdot 2.1338 \end{bmatrix} \cdot 2.1338(0.32292-1) \end{bmatrix} + \\ +2\begin{bmatrix} \begin{bmatrix} 8.9190 \cdot 10^{-6}(0.32292-1) + \\ +1.0741 \cdot 10^{-5}(2.1338-1) \end{bmatrix} \cdot 2.1338 - \\ -\begin{bmatrix} 1.0741 \cdot 10^{-5} \cdot 2.1338 \end{bmatrix}(1-2.1338)(0.32292-1) \end{bmatrix} y_{H_2} + \\ + \begin{bmatrix} \begin{bmatrix} 8.9190 \cdot 10^{-6}(0.32292-1) + \\ +1.0741 \cdot 10^{-5}(2.1338-1) \end{bmatrix}(1-2.1338) + \\ \begin{bmatrix} 8.9190 \cdot 10^{-6}(0.32292-1) + \\ +1.0741 \cdot 10^{-5}(2.1338-1) \end{bmatrix} 2.1338(0.32292-1) - \\ -\begin{bmatrix} 8.9190 \cdot 10^{-6} + 1.0741 \cdot 10^{-5} - \\ -2 \cdot 1.0741 \cdot 10^{-5} \cdot 2.1338 \end{bmatrix}(1-2.1338)(0.32292-1) \end{bmatrix} y_{H_2}^2 \end{bmatrix}}{\left[2.1338 + y_{H_2}(1-2.1338)\right]^2 \left[1 + y_{H_2}(0.32292-1)\right]^2} \Rightarrow$$

$$\frac{d\mu(T)}{dy_{H_2}}(y_{H_2}) = \frac{\left[3.2384 \cdot 10^{-6} - 8.9879 \cdot 10^{-6} y_{H_2} + 4.2656 \cdot 10^{-6} y_{H_2}^2\right]}{\left[2.1338 - 1.1338 y_{H_2}\right]^2 \left[1 - 0.6771 y_{H_2}\right]^2}$$

$$\left\{\frac{d\mu(T)}{dy_{H_2}}(y_{H_2}) \geq 0 \wedge y_{H_2} \in [0,1]\right\} \Leftrightarrow$$

$$\left\{4.2656 \cdot 10^{-6} y_{H_2}^2 - 8.9879 \cdot 10^{-6} y_{H_2} + 3.2384 \cdot 10^{-6} \geq 0 \wedge y_{H_2} \in [0,1]\right\} \Leftrightarrow$$

$$\left\{y_{H_2}^2 - \frac{8.9879}{4.2656} y_{H_2} + \frac{3.2384}{4.2656} \geq 0 \wedge y_{H_2} \in [0,1]\right\} \Leftrightarrow$$

$$\left\{\begin{bmatrix} y_{H_2} \geq \frac{1}{2}\left(\frac{8.9879}{4.2656} + \sqrt{\left(\frac{8.9879}{4.2656}\right)^2 - 4\frac{3.2384}{4.2656}}\right) \vee \\ y_{H_2} \leq \frac{1}{2}\left(\frac{8.9879}{4.2656} - \sqrt{\left(\frac{8.9879}{4.2656}\right)^2 - 4\frac{3.2384}{4.2656}}\right) \end{bmatrix} \wedge y_{H_2} \in [0,1]\right\} \Leftrightarrow$$

$$\left\{\{y_{H_2} \geq 1.6458 \vee y_{H_2} \leq 0.4613\} \wedge y_{H_2} \in [0,1]\right\} \Rightarrow y_{H_2} \in [0, 0.4613] \quad \text{O.E.}\Delta$$